\documentclass[acmsmall,natbib=false,datamodel=acmdatamodel,
style=acmauthoryear]{acmart}

\usepackage{biblatex}

\setcopyright{none}

\begin{document}

\title{Content Moderation by LLM: From Accuracy to Legitimacy}

\author{Tao Huang}
\email{taohuang@cityu.edu.hk}
\affiliation{
    \institution{City University of Hong Kong}
    \city{Hong Kong}
    \country{China}
    }

\begin{abstract}
One trending application of LLM (large language model) is to use it for content moderation in online platforms. Most current studies on this application have focused on the metric of \textit{accuracy} -- the extent to which LLMs make correct decisions about content. This article argues that accuracy is insufficient and misleading because it fails to grasp the distinction between easy cases and hard cases, as well as the inevitable trade-offs in achieving higher accuracy. Closer examination reveals that content moderation is a constitutive part of platform governance, the key of which is to gain and enhance \textit{legitimacy}. Instead of making moderation decisions correct, the chief goal of LLMs is to make them legitimate. In this regard, this article proposes a paradigm shift from the single benchmark of accuracy towards a legitimacy-based framework for evaluating the performance of LLM moderators. The framework suggests that for easy cases, the key is to ensure accuracy, speed, and transparency, while for hard cases, what matters is reasoned justification and user participation. Examined under this framework, LLMs' real potential in moderation is not accuracy improvement. Rather, LLMs can better contribute in four other aspects: to conduct screening of hard cases from easy cases, to provide quality explanations for moderation decisions, to assist human reviewers in getting more contextual information, and to facilitate user participation in a more interactive way. To realize these contributions, this article proposes a workflow for incorporating LLMs into the content moderation system. Using normative theories from law and social sciences to critically assess the new technological application, this article seeks to redefine LLMs' role in content moderation and redirect relevant research in this field.
\end{abstract}

\keywords{Generative AI, LLM, content moderation, platform governance}

\ccsdesc[500]{Human-centered computing~Human computer interaction (HCI)}

\maketitle

\section{Introduction}
LLM (large language model) can be used in a wide range of scenarios. One emergent area of its application is content moderation. Within a span of less than two years, such topic has generated considerable scholarly discussion (see \cite {Francoetal2023}; \cite{Gilardietal2023}; \cite{Huangetal2023}; \cite{Royetal2023}; \cite{Maetal2024}; \cite{Nghiem_Daumé2024}; \cite{Kollaetal2024}; \cite{Vishwamitraetal2024}; \cite{Kumaretal2024}; \cite{Thomasetal2024}). Most studies so far have deemed LLM promising in this application field, and the technology has been hailed as “the greatest change to the dynamics of content moderation in at least a decade” (\cite{Willner_Chakrabarti2024}).

The rationale behind the promising tone is that the new tool, through appropriate design and use, could be more accurate in detecting and classifying violating content than its predecessors. In other words, the existing scholarship on LLM moderation has centered upon the metric of accuracy – the extent to which LLM makes correct decisions of moderation. LLMs' superiority in accuracy derives from their technical structure and pretraining process, which enable them to better understand complex contexts as well as to adapt to changing circumstances. Such accuracy bonus drives the advocacy of many researchers that the new technique should be encouraged for wider use in managing content (\cite{Wengetal2023}).

This article argues, however, that accuracy is far from enough in situating and evaluating LLMs' role in content moderation. On the one hand, it is impossible to make all decisions right. Content moderation involves inevitable tradeoffs and balances among various rights and interests. On global platforms, in particular, people from different cultural backgrounds may reasonably disagree on how the balance or tradeoff should be made. Like judicial adjudication by courts, moderation by online platforms also faces hard cases. On the other hand, accuracy is not the only goal of a moderation system. Content moderation is not an isolated process but a constitutive component of the platforms’ governance scheme. The central purpose of such a scheme is to provide legitimacy to the private exercise of governance power, which, unlike state governance, has neither constitutional authorization nor democratic approval. To justify their power of delimiting people’s fundamental freedoms online, platforms must maintain a governance scheme that gains legitimacy for them. As a vital component of governance, content moderation must also serve this central goal, rather than the narrow focus of improving accuracy.

This article proposes a paradigm shift from the \textit{accuracy} discourse to a more comprehensive analytical framework based on \textit{legitimacy}. That being said, the right question to be asked is not whether and how LLMs can help reach more accurate moderation decisions, but whether and how it can enhance the legitimacy of the moderation and governance system of online platforms. Legitimacy is a broad concept that consists of far more factors than accuracy: it examines how the practice of moderation could be normatively justified and accepted. The framework proposed by this article distinguishes between easy cases and hard cases and uses different legitimacy metrics to measure the moderation of the two categories. For the easy cases, which are those containing clear answers and little controversies, moderating them should be accurate, fast, and transparent. When straightforward answers are available, the moderation system should deliver those answers in an efficient and open manner, ensuring the health and vividity of the online discourse. For hard cases, which involve complex fact contexts or difficult value compromises, requiring fast and correct decisions is unrealistic, since there may hardly be any agreement on what decision is correct. Rather, the legitimacy of deciding hard cases depends upon the substantive quality of the explanation (reason-giving) for the decisions, as well as the procedural justice of the decision process (especially the participation of users). 

Examining LLMs under this framework reveals that the most promising strength of this tool is not in increasing the accuracy of moderation. LLMs' accuracy advantage is significantly discounted by their increased latency and cost in moderating easy cases, as well as the diminished relevance of accuracy in hard cases. Instead, LLMs can play valuable roles in various other aspects that promote the legitimacy of platform governance, such as to conduct pre-screening of hard cases from easy cases, to provide quality explanations for moderation decisions, to assist human reviewers in getting more contextual information, and to facilitate user participation in a more interactive way.

The accuracy discourse is parochial, misleading, and counter-productive in the sense that it has led substantial academic and industrial resources to an enterprise that produces little return. Achieving higher accuracy is not the best application scenario here; researchers and developers interested in LLM moderation should consider reorienting their focus of efforts. Locating LLMs within the legitimacy framework helps the reorientation by aligning the technological project of progressive improvement with the socio-legal concerns of normativity and justification. It allows us to rethink LLMs' real potential in content moderation and platform governance, as well as what improvements can be made to further realize its potential.

\section{The Accuracy Discourse}

Researchers and developers have a growing interest in using LLMs to moderate content on digital platforms and communities. Most existing studies on LLM moderation have focused on the metric of accuracy. This metric is usually measured by the percentage of correct decisions (true positives plus true negatives) among the total number of cases. There are metrics that are akin to accuracy, such as recall and precision, and there might be differences and tensions among them (\cite{Gonganeetal2022}, 21; \cite{Sartor_Loreggia2020}, 45). However, the general goal of all AI moderation tools is to reduce the two types of errors and increase the ratio of correct ones. This article uses accuracy in this broad sense, referring to the capability of making correct (true) decisions and avoiding erroneous (false) ones.

This article uses "the accuracy discourse" to describe the dominance of the accuracy metric in the current literature on moderation by LLM. Such discourse can be demonstrated by several sources. First, Among the 1220 results in Google Scholar after a full-text search of keywords “LLM” and “content moderation,” 844 of them contain “accuracy,” constituting roughly two-thirds. (Note: The search was conducted on Aug 5, 2024.) Second, the author of this article has manually reviewed those articles and found that most papers discussing moderation by LLM focused on the metric of accuracy. Indeed, all the papers on LLM moderation cited by this article have explored LLMs' accuracy performance. The list could go on during the process of writing this article (for example, \cite{nr_Dingetal2024}, \cite{nr_Zhaoetal2024}, \cite{nr_Pérezetal2024}, \cite{nr_Kang2024}, \cite{nr_Ouyangetal2024}, \cite{nr_Sasidaran_J2024}, \cite{nr_Xuetal2024}, \cite{nr_Antypasetal2024}). Third, the accuracy discourse has also been corroborated by \cite{nr_Penagos2024}'s and \cite{nr_Gielensetal2025}'s review of the literature.

Generally, studies to date reported positive results regarding LLMs' accuracy performance in moderation tasks. (\cite{Gilardietal2023}, 2) found that in most cases, ChatGPT’s zero-shot accuracy is higher than that of MTurk (crowd-workers of annotation on Amazon). (\cite{Ottosson2023}, 9) attested that “LLMs can perform on par or better than traditional ML models in cyberbullying detection. (\cite{Kokaetal2024}) revealed that LLMs exhibit a high degree of accuracy in detecting fake news. (\cite{Choi_Ferrara2024}) tested the fine-tuned LLMs’ performance in fact-checking and discovered that “LLMs can reliably assess the relationships between social media posts and verified claims, offering performance comparable to human evaluations”. Similar work of testing LLMs' ability in fact-checking has been done by (\cite{Quelle_Bovet2024}). What’s more, (\cite{Vishwamitraetal2024}) has used chain-of-thought reasoning to prompt the LLMs and observed a superior rate of accuracy in detecting new waves of online hatred; and \cite{nr_Wangetal2024} has reported LLMs can achieve approximately 95\% accuracy in moderating images.

Using LLMs to moderate online content not only attracted much scholarly attention but also generated heated discussion in the industry. As early as 2020, Meta has tried to incorporate LLMs into its toolkit for detecting hate speech(\cite{nr_Meta2020}). Recently, OpenAI has enthusiastically touted its product, GPT4, for conducting content moderation tasks (\cite{Wengetal2023}; \cite{Fried2023}). The chief reason for OpenAI’s enthusiasm is GPT4’s impressive F1 score (\cite{Wengetal2023}) -- the harmonic mean of precision and recall (\cite{Akre2023}). Even though current models are far from perfect, researchers believe that they can be continuously improved to achieve greater performance in accurate moderation (\cite{nr_Gielensetal2025}, 16).

To be sure, not all authors are optimistic. Some also explored LLMs' limitation in accurate moderation, at least at the current stage (\cite{Boicel2024}; \cite{Kumaretal2024}). Particularly, current LLMs still rely substantially upon fine-tuning for moderation: while fine-tuned LLMs achieve impressive accuracy scores (\cite{nr_Roumeliotisetal2024}), off-the-shelf LLMs without fine-tuning are far less competent, especially in moderating contextual content such as hate speech (\cite{Kollaetal2024}, \cite{Shethetal2024}, \cite{nr_Masudetal2024}, \cite{Royetal2023}). 

In any case, there is no doubt that accuracy has been used as the dominant benchmark in the scholarship of LLM moderation. Accuracy is surely important, and it is natural or intuitive to expect moderation decisions to be correct. However, the accuracy discourse is parochial and misleading. It attracts most scholarly attention and resources in this field, some of which should have been spent on other equally if not more, important issues. Moving our interests and endeavors beyond the accuracy discourse would significantly enrich our research agenda and make our enterprise more productive. Before a full critique of the accuracy discourse in part 4, the next part introduces some unique features of LLMs that may explain their accuracy performance. This can help unearth the rationale behind the optimistic tone of the accuracy discourse, as well as provide some necessary technical backgrounds of the new tool.

\section{The Unique Features of LLM Moderation}

AI moderation tools have been widely used by online platforms and communities to manage content, as human moderation is impossible to scale and costly to operate. Currently, the most widely used is the technique of machine learning (ML). Since LLM is one sub-category of ML, this article uses “traditional ML” to refer to the technique of ML before the emergence of LLM. Traditional ML moderation contains several steps, including the development of the model, the annotation of the training dataset, the training of the model by the annotated dataset, and the use of the model to moderate real-life content (\cite{DAlonzo2023}); \cite{Sartor_Loreggia2020}). The key condition for this technique is human supervision, which determines the quality of the annotated datasets as well as the learning process (\cite{Maetal2024}, 2).

The traditional ML method for content moderation has several limitations: First, it heavily relies on manual annotation of the training dataset. Such practice generates substantial costs for subsidizing the labor (\cite{Wangetal2023}; \cite{Lietal2024}, 2) and also introduces biases into the process (\cite{Dentonetal2021}). In addition, different human annotators may not agree on the decision of a piece of content, especially for highly subjective categories such as hate speech (\cite{Gonganeetal2022}, 35, 10; \cite{Llansóetal2020}, 8; \cite{Marsoofetal2023}, 79). Second, the traditional method lacks flexibility and adaptability. A model trained by one dataset can hardly perform equally well in other datasets, represented by different cultures, languages, or contexts (\cite{Marsoofetal2023}, 78; \cite{nr_Maliketal2024}). Adapting to changing circumstances is difficult because doing so would require re-annotation of the dataset and re-training of the model (\cite{Mullicketal2023}, 562).

LLMs can perform better in these two aspects. Technically speaking, the major difference between the traditional ML models and LLMs is that the former are target-trained with specific datasets, while the latter are pre-trained with a massive corpus of online data. LLMs are based on deep learning (DL) techniques which can “extract pertinent features from textual data, eliminating the requirement of manual feature engineering – a typical practice in conventional ML approaches” (\cite{Rawatetal2024}, 34). In other words, LLMs are self-supervised (\cite{Surden2024}, 1959), as they can automatically learn implicit language patterns from the immense body of online materials without human annotation and supervision. This key difference explains its potential to overcome the two limits of traditional models and achieve better accuracy scores.

First, LLMs have the potential to better understand contexts and nuances. There are two reasons for this. One reason is the distinctive training process of LLMs. The pretraining of LLMs by a large corpus of data exposes the models to a wide range of content from diverse sources (\cite{Bhattacharyaetal2024}, 16-17). Those sources may contain billions of web documents, potentially covering most areas of knowledge that have been stored online (\cite{Surden2024}, 1961). Such scale and diversity enable LLMs to generalize across different domains and to develop a comprehensive understanding of common language use (\cite{nr_Latifetal2025}). Broader exposure empowers LLMs with the capacity to capture more contextual expressions and idiomatic phrases, such as irony, sentiment, and sarcasm. Another reason why LLMs excel in contextual reasoning lies in the transformer architecture used by most mainstream LLMs (including the popular GPT) (\cite{Menon2023}). Such architecture uses the mechanism of self-attention, which takes the whole content of user input into consideration, even when the distance between two words is remote (\cite{Surden2024}, 1964). In other words, it weighs the importance of each word in a sentence in relation to \textit{every} other word, rather than the neighboring words only (\cite{Vaswanietal2023}). Thus, while traditional ML models struggle to understand complex and non-linear correlations, LLMs hold the capacity to capture long-range dependencies and contextual relationships within user input rather than grasping meanings through its fragmented parts. The contextual analysis capability of LLMs, acquired from their training and structure, gives them impressive potential in content moderation (\cite{nr_AlDahouletal2024}).

Second, LLMs are more adaptable and flexible than traditional ML models. LLMs do not rely on manual labeling of datasets; rather, their pretraining is unsupervised (\cite{Divakaran_Peddinti2024}, 1). Because the decision logic is decoupled from the model, LLMs can quickly adapt to changes in policy and context without re-annotating the data and re-training the model (\cite{Mullicketal2023}, 562). Through pretraining, LLMs already acquire contextual knowledge and understanding across a diverse range of areas; when new context emerges, LLMs can be adjusted to fit with the new context through prompting or fine-tuning (\cite{Ottosson2023}, 3; \cite{Gületal2024}, 2; \cite{nr_Mathewetal2021}). Such adaptability renders LLMs especially capable of dealing with emergent crises or events, such as new waves of hatred online (\cite{Vishwamitraetal2024}, 2). To be sure, the literature is not consensual on this topic: there are also works challenging LLMs' generalizability strength (\cite{Shethetal2024}, \cite{nr_YizhuoZhangetal2024}), suggesting the need for more exploration. LLMs' generalizability or adaptability also has temporal limits since they will decrease over time (\cite{nr_ChenghaoZhuetal2024}). In this sense, although LLMs may be more adaptable than traditional ML models due to its larger knowledge base, its knowledge still needs to be updated periodically.

It should be noted that apart from those strengths, LLMs also contain inherent limits or risks. The two most typical ones are hallucination and bias. LLMs can generate hallucinations (\cite{nr_ZiweiXuetal2024}), referring to the discrepancies of their responses with either real-world facts or user input (\cite{nr_LeiHuangetal2025}). Bias can also be reflected and amplified by LLMs, reinforcing social stereotypes and inequalities (\cite{nr_Gallegosetal2024}). Despite these limitations, LLMs' capability in the knowledge base and contextual understanding has brought about an optimistic tone in the accuracy discourse. Researchers are working enthusiastically on improving LLMs' accuracy performance.

\section{Why Accuracy is NOT Enough}

The accuracy discourse is parochial and misleading. Taking accuracy as our primary or even exclusive metric for evaluating and designing LLM moderation tools misunderstands the function of accuracy and its position in the governance system. There are four cases against the accuracy discourse. First, it is impossible to achieve perfect accuracy in practice, and trying to do so is dangerous. Second, accuracy has both individual and systematic aspects, but the current discourse stresses the former while ignoring the latter. Third, the discourse failed to recognize the distinction between easy cases and hard cases, which should be taken differently in platform governance. And fourth, focusing exclusively on the substantive result of decisions overlooks other aspects of content moderation that are also critical for governing the platforms. This part will illustrate these four points in turn.

\subsection{\textit{Can we make all decisions right?}}

The accuracy discourse has one assumption behind it: that there exists, and we can identify, an objective or consensually endorsed standard of determining the right decisions (true positives and true negatives) from errors (false positives and false negatives). That assumption, however, does not stand. Determining whether certain content violates the platform’s rules involves probing the facts, identifying the relevant rules, and making a choice in front of value conflicts. It requires delicate assessment and shrewd judgment. Without objective standards, the only metric for measuring algorithmic decision-making is the decision of humans. Actually, human moderators’ classification has been used as the “ground truth” that trains and evaluates AI tools (\cite{Sartor_Loreggia2020}, 46). Facebook has hired a separate reviewing panel (containing three reviewers) to evaluate the correctness of first-instance human moderators (\cite{Bradfordetal2019}, 13). This is also the case for experiments that test LLMs' capability of moderation (\cite{Huangetal2023}, 3).

The problem with this approach is that human moderators sometimes differ in moderating cases. “By definition, content moderation is always going to rely on judgment calls, and many of the judgment calls will end up in gray areas where lots of people’s opinions may differ greatly” \cite{Masnick2019}. One salient issue, as noted by many researchers in this field, is the intercoder disagreement, referring to the instance that different moderators cannot reach a consensus on a case (\cite{OConnor_Joffe2020}). Even though intercoder disagreement can be addressed by some methods of finding unified solutions, such as the Kappa score (\cite{nr_Gwet2014}), there are cases that are inherently controversial, and consensus is impossible to achieve. In such circumstances, whose decision should be counted as “ground truth” or the standard of accuracy?

The existence of disagreement reveals that we cannot make every case right, because not every case can be decided by humans in a consensual and non-controversial way. In fact, “[n]o legal system guarantees a right to an accurate decision. Enforcement of speech rules has never been perfect, online or off” (\cite{Douek2021}, 799). People reasonably disagree on many issues, especially issues regarding moral, religious, and other value judgments (\cite{Waldron1999}). Observation on legal adjudications shows that determinacy and consensus are not always available. Just take a look at the free speech cases decided by the U.S. Supreme Court: within the past 20 years, the court decided 43 cases on free speech; 26 of them contain dissenting opinions (about two-thirds!), and 7 of them are decided with a divisive vote of 5:4. (Note: The statistics here is up to the date of June 18, 2024.) The prevalent disagreement on legal cases illustrates that consensus on accuracy is an illusion.

The goal of perfect accuracy is not only impossible to achieve, it is also undesirable. Perfect accuracy means there will be, and must necessarily be, a single, uniform, and perfect standard for measuring accuracy. If there is not, we must design one. However, doing so would greatly suppress the space of value contestation by imposing one set of values upon the whole platform. Consider the fact that most online platforms are global, with users coming from divergent cultural backgrounds (\cite{Bradford2020}). Striving for uniform and definite applications of rules may serve the goal of formal consistency, but sacrifices other vital principles such as value pluralism, open disagreement, and epistemic humility. This is unacceptable for governing global and open platforms.

\subsection{\textit{Two levels of accuracy}}

Current studies on LLM moderation use accuracy at the individual case level, aiming to make correct decisions in each and every individual case. This perspective ignores that accuracy can also be measured at the holistic or system level. Inspired by the idea of "systems thinking" in content moderation (\cite{Douek2022}), I raise the notion of "system accuracy" here. System accuracy is not the aggregate of the accuracy scores of individual cases. Rather, it refers to the general performance of the whole moderation system, including metrics like consistency, predictability, and the fairness of the distribution of errors. Three insights can be gained by taking the system level of accuracy into account.

First, accuracy itself contains competing goals that call for tradeoffs and balance. Researchers generally use precision and recall to measure the two different aspects of accuracy performance. Precision measures a model's capability to reduce false positives, while recall measures its ability to avoid false negatives (\cite{nr_Google2024}). There usually exists a tension between the two metrics: prioritizing one would compromise the other (\cite{nr_Pérezetal2024}). When perfect accuracy cannot be achieved, we must make tradeoffs. The choice between false positives and false negatives is a value choice of whether we assign more importance to combating more bad speech and ensuring healthy communications online or allowing more benign speech and promoting free and inclusive expression. Even though precision and recall could measure the outcome of the value tradeoff, the two metrics cannot reflect the process of making the tradeoff or the justifications or considerations behind it. Instead, user control and participation are necessary for justifying the tradeoff (\cite{nr_Caoetal2024}).

Second, the perspective from the system level further corroborates why perfect accuracy is not a worthwhile goal. One can reasonably anticipate the moderation of one case to be accurate. But things will get much harder if the number of cases arises to millions or billions, and the time for moderating each case is very limited (sometimes only seconds for a post) (\cite{Douek2022}, 549). Moderation also takes costs, no matter who does the job – human or AI. The factors of time, cost, and scale at the system level render moderation a managerial enterprise that must constantly make tradeoffs under limited resources. Errors are to be managed by considering costs, not to be avoided at all costs.

Third, viewed through the system lens, accuracy is not only a numerical metric, measuring the performance of moderation by percentages. Rather, accuracy is also a distributional metric. If, as the foregoing analysis suggests, errors are inevitable and perfect accuracy is illusory, then how to manage accuracy and how to distribute errors becomes crucial (\cite{Douek2021}, 791). The right questions worth examination should include, for example, “[w]ho suffers from false positives and false negatives? ... [w]hich types of errors are known and tolerated, [and] how is risk distributed” (\cite{Ananny2019}). Accuracy at the individual case level is a poor indicator of these issues since it ignores the distributional effect. Even though the total accuracy is sufficiently high, say 99.99\%, it would still be worrisome if the tiny percentage of errors (0.01\%) were mostly suffered by LGBTQ groups or racial minorities. 

\subsection{\textit{Easy cases vs. hard cases}}

The accuracy discourse ignores a key distinction: that between easy cases and hard cases. Distinguishing between easy cases and hard cases is reasonable and even necessary for three reasons. First, such distinction already exists in platforms’ content moderation practice. Meta has reported that “[i]n most cases, identification is a simple matter. The post either clearly violates our policies or it doesn’t. Other times, identification is more difficult” (\cite{Meta2022c}). This claim not only acknowledges the distinction between easy and hard cases but also discloses the fact that easy cases count for a dominant majority of all moderation cases, while hard cases are relatively few. Platforms constantly draw the line between easy and hard cases when they divide labor between AI moderators and human moderators (\cite{Bradfordetal2019}, 12, 14). For example, Meta uses AI tools to “identify and remove a large amount of violating content—often, before anyone sees it”, and when “technology misses something or needs more input”, human reviewers will intervene to make their judgment calls (\cite{Meta2022b}). Such practice reveals that platforms must make, and already have made, the strategic choice of treating different cases differently.

Second, easy cases and hard cases have also been distinguished in legal systems, for similar reasons (see generally \cite{Hart1958}; \cite{Dworkin1975}). In most instances, we get highly determinate answers regarding what the law prescribes (\cite{Schauer1985}, 423): consider our behaviors adjusted according to the law, the quick and clear advice we receive from lawyers during consultation, and cases that are settled or finalized at trial level (\cite{Schauer1985}, 412-3; \cite{Fischman2021}, 597; \cite{Watson2023}, 225). Sometimes, the result of a case is unclear or contested, so value judgments are needed (\cite{Fischman2021}, 596): think about the cases that are appealed or those with dissenting opinions from the deciding judges. In some sense, the institutional design of litigation costs, hierarchies of courts, and judicial writing of dissents are made to distinguish between easy cases and hard cases so that limited judicial resources can be reasonably distributed and clear issues can be quickly resolved while difficult ones can be sufficiently debated and contested.

Third, easy cases and hard cases have different social impacts and different expectations from the users. For easy cases, answers are relatively clear and obvious; what users and society expect is their resolution in a correct and convenient way. Substantive results are what matter the most, so accuracy should be the dominant metric here. Hard cases, by contrast, are controversial and contested. They typically involve complex facts or contentious value judgments, and people do not and may never have consensual answers to those cases. Should images of women's breasts and nipples be prohibited (\cite{Demopoulos2023})? If there are exceptions, what exceptions should there be? If child nudity is generally banned but can be allowed in some cases due to, e.g., newsworthiness or awareness raising, how should the boundaries be drawn (\cite{Kleinman2016})? What about hate speech like Holocaust denial (\cite{Dwoskin2020})? When reasonable people differ on the substantive results, what matters most is not who wins and who loses, but how the result has been reached.

\subsection{\textit{Content moderation is a part of governance}}

The accuracy discourse views content moderation as an isolated process, a process in which platform rules are mechanically applied to individual cases. Framed in this way, content moderation is like a syllogistic game. But content moderation is never simply the application of rules or the adjudication of disputes. Rather, it is, and must be taken as, a constitutive part of the governance system of platforms. The personnel, budget, and tools for content moderation are all internal components of the governance structure of platforms (\cite{Gillespie2018}). Machine tools like LLMs should be integrated into the systematic effort, together with humans, in the governance of the online space (\cite{nr_Ruckenstein_Turunen2020}).

For the governance of those platforms, the central concern should be legitimacy – the concern of how platforms’ exercise of power could be justified and recognized (\cite{Suzoretal2018}, 387; \cite{Bloch-Wehba2019}, 68; \cite{Freeman2000}, 666). This is due to two, interrelated reasons: rights protection and power check. Moderation is the process of delineating the boundaries of free speech when it conflicts with other rights or interests. The exercise of many basic rights of people is dependent on private platforms (\cite{Klonick2018}, 1668). Public and human rights law prescribes that any state intervention or infringement on fundamental rights must be appropriately justified to be legitimate. The stakes are not lower when the infringing body is a private company. These private platforms exert tremendous power to influence people’s basic freedoms. As (\cite{Rasoetal2018}, 38) summarized, “the largest online platforms, such as Facebook and Google, exercise more power over our right to free expression than any court, king, or president ever has—in view of the very significant percentage of human discourse that occurs within the boundaries of these ‘walled gardens.’” The power of delimiting people’s fundamental rights must be legitimate for it to earn acceptance, obedience, and respect (\cite{Solum2004}, 178).

As an essential and definitional component of online platforms (\cite{Gillespie2018}, 21), content moderation must be viewed as a constitutive unit of governance, the goal of which is to legitimize the exercise of platforms’ power of defining people’s fundamental rights and shaping the contour of the online public square. This article argues for a paradigm shift from accuracy to legitimacy as the guiding principle for content moderation and platform governance. The next part introduces a legitimacy-based framework.

\section{A Legitimacy-Based Framework for Content Moderation}

\subsection{\textit{Introductory notes on the framework}}

Even though previous works have proposed frameworks for normatively guiding the content moderation practice (\cite{Marsoofetal2023}, \cite{nr_Ruckenstein_Turunen2020}, \cite{nr_Nahmias_Perel2021}), they fell short of comprehensiveness and delicacy. Elaborating on existing efforts, my framework covers both substantive and procedural aspects of legitimacy, as well as distinguishes between easy cases and hard cases. To be sure, my framework is far from perfect - it serves as an invitation for further discussions and endeavors.

Legitimacy has both substantive and procedural aspects (\cite{Bagley2019}, 379). Substantive legitimacy evaluates the content of decisions – whether they are correct, fair, or conforming to some high values or principles (\cite{Mondak1994}, 676-7). Another crucial component of legitimacy is procedural justice, which refers to the approach, manner, and process in which rules are enacted and enforced \cite{Tyleretal2015}. Justice Marshall has succinctly summarized the two aspects as “the prevention of unjustified or mistaken deprivations and the promotion of participation and dialogue by affected individuals” (\cite{Marshall1980}). This is also the case for content moderation. Users care about both the outcome of their cases and how they have been treated in the process. The public holds serious concern over how the platforms’ moderation system performs as well as to what extent it remains accountable.

Easy cases and hard cases should also be differentiated under the framework. Separating the two categories might be a challenging task (\cite{Gillespie2020}, 3). Meta has described the hard cases as involving content that is “severe, viral, nuanced, novel and complex” (\cite{Meta2022c}). But these descriptions are too abstract to be manageable. Instead of offering a bright line, I argue that a case would become hard out of one of the following conditions:

1) Complexity of facts or contexts. Disagreement about facts is one type of cause that makes a case hard (\cite{Hutchinson_Wakefield1982}, 93). Lack of clear and comprehensive understanding of the contextual facts of a case may inhibit judges or moderators from reaching a clear answer. One pertinent factor here is the category of content. Some content generally requires less contextual knowledge, such as spam, child porn, and IP-infringing materials. Other categories of content may be highly contextual and culturally dependent, such as hate speech, which is more likely to make a case hard.

\textit{Example:} In a work that examines LLMs' performance under task determinacy, \cite{nr_Guerdanetal2024} has raised an illustrative example in which a hard case arises out of insufficient information of context. The moderation task is to determine whether the statement "William is such a Cheesehead" is derogatory. As the authors analyzed, "Whereas an American rater might view 'Cheesehead' as an endearing reference to a Green Bay Packers Football fan, and respond No, a Dutch rater might connect 'Cheesehead' to its historical use as a WW2-era pejorative slur, and thus respond Yes." (\cite{nr_Guerdanetal2024}) To judge the appropriateness of the content, more contextual information is needed: such as the audience of the speech and the intent of the speaker. In this case, the complexity of facts and contexts makes it hard.

2) Vagueness of rules. Sometimes the moderation rules may be too vague to dictate a clear or singular result. The ambiguity of rules is more complicated than the non-existence of rules. When there are no rules applicable to a case, we can resort to the principle that those that are not prohibited are permitted (\cite{Soper1977}, 484-5). But when a rule is so ambiguous in its applicability or that it dictates multiple reasonable but potentially conflicting results (\cite{Hutchinson_Wakefield1982}, 94), moderators have to use other tools, sometimes judgment calls, to pick from those conflicting options.

\textit{Example:} In one case reviewed by the Meta Oversight Board, experts of the Board ruled that Meta's content rules are too vague. The case is about a user pose that was incorrectly attributed to Joseph Goebbels and was removed by Meta for violating its community rule on Dangerous Individuals and Organizations. The Meta rule says it will "remove content that expresses support or praise for groups, leaders or individuals involved in these activities"; the Board, however, pointed out that the rule does not contain any examples explaining what is "support" or "praise", and it also fails to provide a list of individuals and organizations designated as dangerous (\cite{FacebookOversightBoard2020}).

3) Plurality of rules. This arises when there are multiple rules that are applicable to a case (\cite{Hutchinson_Wakefield1982}, 94; \cite{Schauer1985}, 415). Sometimes these different sources of rules may be in conflict. This is most salient in the case of value conflict when different values are inherent in different rules. On these occasions, decision-makers must delicately choose from conflicting rules or values.

\textit{Example:} In the famous case of the “Napalm Girl” photo, a picture showing a nude girl fleeing a Napalm attack during the Vietnam War has been removed by Facebook out of its policy prohibiting child nudity; later, the platform decided to restore the content due to its journalistic and historical importance (\cite{nr_Levinetal2016}). This case has generated huge controversies in the media. As a hard case, it reflects the tension between two different rules of the platform as well as two different values behind: prohibiting child nudity and protecting children's safety on the one hand, and allowing awareness-raising content and promoting free expression on public matters on the other (\cite{Gillespie2018}, 1-5).

4) The textual meaning of a rule contradicts its underlying purpose or value or violates some established principles of morals (\cite{Schauer1985}, 415-6; \cite{Hutchinson1989}, 560). Sometimes the application of a rule may lead to unconventional or controversial results (\cite{Soper1977}, 488). This requires the decisionmaker to make a choice between adhering to the textual mandate and resorting to some external values or morals.

\textit{Example:} The Meta Oversight Board, for example, has three sources of rules for its judgment: Meta’s Community Standards, Meta’s values, and the International Human Rights Law (IHRL) (see \cite{MetaOversightBoard2022}).  Sometimes the Community Standards may conflict with higher rules (such as IHRL). This happened in one case, in which a user's comment calling another user "cowardly bot" was removed by Facebook because the word "cowardly" referred to a negative character. In reviewing the case, the Oversight Board held that Facebook’s moderation decision conforms with its community standards, but the decision should be overturned because it violates the Values of Facebook as well as the IHRL (\cite{FacebookOversightBoard2021}).

Below, I introduce the framework based on the core benchmark of legitimacy and the distinction between easy and hard cases. As a general framework for evaluating content moderation, it is not limited to applying to LLMs. Rather, the framework can be used to measure the performance of different moderators (e.g. traditional ML, LLM, ordinary human moderator, and expert moderator) according to the legitimacy criterion.

\subsection{\textit{Easy cases: making correct, fast, and transparent decisions}}

Easy cases are those with clear answers: either the post has violated platform rules or not, and the type of violation is obvious. Moderating this type of cases is the routine job of moderators. From the legitimacy perspective, moderating easy cases must serve two goals: for users, cases should be resolved in a correct and timely manner; for platforms, the communicative space must be regulated fairly, efficiently, and openly. These are the goals that are not only achievable but also indispensable for gaining trust for the moderation practice. From these goals, we can derive three legitimacy metrics for moderating easy cases:

1) \textit{Accuracy}. Users expect cases to be decided accurately, and platforms also anticipate the right moderation in order to gain public trust and build amicable communities. Accuracy should be measured at both individual-case level and system level. The former refers to the aggregate percentage of correctly decided cases, while the latter requires the distribution of accuracy -- or, viewed from the opposite angle, errors -- to be fair across different groups of people, categories of content, and periods of time.

2) \textit{Speed}. One salient difference between online moderation and offline speech adjudication is the velocity and virality of communications in the former context. In the online world, it is more appropriate to say “justice delayed is justice denied”. For cases with obvious answers of right or wrong, providing the right answers in a timely manner means delivering justice to the users. Speed is also crucial for the moderation system as a whole, since maintaining the health of the platform requires fast removal of harmful content.

3) \textit{Transparency}. Transparency refers to the extent to which the moderation process is visible to the public. This metric ensures that the users are informed about the moderation decisions as well as the internal moderation process (\cite{Suzoretal2019}, 1527). As an old principle of justice, transparency can be imposed on the individual case level and the system level. The Santa Clara Principles, for instance, urge the platforms to not only disclose the numbers (statistics) of the moderation system but also provide users with individual notices about the moderation decisions (\cite{AccessNowetal2021}). To be sure, the transparency requirement at both levels can be either loose or strict, subject to considerations of practical needs and costs.

\subsection{\textit{Hard cases: enabling justification and participation}}

Hard cases defy clear and consensual answers. Even though the hard cases only constitute a tip of the giant moderation iceberg, these “high-profile content moderation controversies” act as paradigms that shape the public discourse and opinions about platform governance. For the users, especially the losing party, to accept and respect the decisions of hard cases, the key is not the substantive result, but the process of reaching that result. On the one hand, contentious decisions must be justified to be accepted; on the other hand, users and stakeholders must be fairly treated in the moderation process. The two principles have been endorsed by the EU's Digital Services Act: Article 17 and Article 20 prescribe the right to explanation and the right to appeal respectively (\cite{nr_EuropeanUnion2022}).

1) \textit{Justification}. Justification is one of the most important factors that affects users' perception of the fairness of moderation (\cite{nr_Jhaveretal2019}). The more contentious the case, the more it needs to be justified. Jurists like Hart and Dworkin shared the view that for hard cases, adjudicators need to justify their decisions (\cite{Crowe2019}, 80). The metric of justification mandates that explanations be provided to the relevant parties as well as the public. This echoes with the principle of reason-giving in public law (\cite{Bloch-Wehba2019}, 75). Explanation here is not the same thing as the metric of transparency in easy cases: while the latter refers to the disclosure of moderation details in a systematic, statistical, and holistic manner, the former evaluates the quality of explanation in individual cases. Explanation has been recognized as an important factor of legitimacy: for example, empirical findings show that “users who did receive some sort of explanation from moderators regarding their removal were less likely to have posts removed in the future” (\cite{Katsarosetal2024}, 71). In addition, reading explanations of moderation decisions is an educational opportunity for users to learn and internalize the community norms (\cite{Jhaveretal2019}).

To be sure, explanations should not aim to gain wide approval on the merit of the decisions. But refraining from the ambition of achieving consensus on substantive results does not mean that all the substantive reasonings are equally persuasive and equally acceptable. Not all explanations can be qualified as justifications (\cite{Brennan-Marquez2017}, 1288). Justificatory effect entails that the explanations provided must reach a certain level of substantive quality. For example, how the decisions address the facts and rules of the case as well as the context of the controversy, how the decisions respond to users’ concerns and public expectations, and how the decisions approach pressing issues like the borrowing of human rights norms into private moderation context, are all important aspects for measuring the substantive legitimacy of the explanations.

2) \textit{Participation}. If justification (reason-giving) earns substantive legitimacy for moderating hard cases, then participation secures the procedural side of legitimacy. That participation is a central tenet of procedural justice is “as old as the law” (\cite{Solum2004}, 308). Scholars found that people would deem algorithms more acceptable when they are more informed about how the algorithms work and afforded more control in their work (\cite{Moketal2023}, 4, 20). This corresponds with the general finding that “[a]ffected individuals are more likely to perceive a decisional system as legitimate ‘when they play a meaningful role in the process’” (\cite{Kaminski2019}, 1548).

Importing principles from public and constitutional law, the participation metric prescribes that users should have a fair say in the moderation process: this includes, e.g., the right to be informed or noticed, the right to comment, and the right to appeal (\cite{Suzor2018}, 7-8; \cite{Suzoretal2018}, 391-2). The normative meaning for hard cases is not their perfect resolution, but the process of debating and deliberating them in a public way -- that is crucial and even constitutive for an open society (\cite{Gillespie2020}, 3-4). The procedural safeguards ensure that hard cases can be contested in an accessible and inclusive manner. Of course, these procedural requirements are also a matter of degree, subject to practical limits of cost and other considerations.

The defining feature of this analytical framework is that it makes three kinds of distinctions: that among different types of cases, that among different categories of moderators, and that among different specific metrics. In this regard, to examine the performance of LLMs under the framework, we need to ask the following questions: will LLMs be used to moderate easy or hard cases, what are LLMs' strengths and limits on the relevant metrics, and compared to whom.

\section{LLM Moderation under the Framework}

This part argues that accuracy is not LLMs' major field of contribution to content moderation. For easy cases, replacing traditional ML models with LLMs may bring some accuracy bonus, but LLMs also generate additional cost and latency; and as easy cases do not involve much contextual complexity, traditional ML models already perform reasonably well in this category. For hard cases, accuracy is not an important metric, since the standard of what is accurate becomes blurred here. Rather, what matters is the quality of justification for the decisions as well as the procedural justice offered by the decision-making process. 
 
\subsection{\textit{Reassessing LLMs' accuracy capability}}

Examining LLMs under the framework proposed by this article and comparing their performance with other types of moderators, we can see that LLMs' superior capability in accuracy comes with significant limitations and costs.

First, even though LLMs have achieved impressive improvement in accuracy, they cannot take the place of human experts in moderating hard cases. On the one hand, LLMs' accuracy diminishes significantly when the content's level of nuances or implicitness increases (\cite{nr_YuxinWangetal2024}). On the other hand, even though some studies indicated that LLMs' accuracy score is higher than the outsourced human moderators (\cite{Gilardietal2023}, 2; \cite{nr_Joetal2024}, 18), LLMs cannot, at least for the short run, reach the level of experts. According to OpenAI, LLMs and ordinary human moderators (with light training) perform equally well in labeling accuracy, but both are outperformed by expert moderators who are well-trained (\cite{Wengetal2023}). Expert moderators, with systematic training, sufficient decision time, and organizational and financial support from the platforms, are the best performers in accuracy – actually, expert labeling has commonly been used as the “ground truth” for determining what is accurate (\cite{Bradfordetal2019}, 13). To be sure, it is impractical for experts to moderate \textit{easy} cases due to scale and budgetary limits. But the quality of their reasoning and justification makes experts the best choice for moderating \textit{hard} cases. That’s why Meta has delegated some of its hard cases to a board of experts (Oversight Board), the role of which can never be fully replaced by LLMs.

Second, if LLMs' accuracy superiority cannot find its place in moderating hard cases, how about the easy cases? In the realm of easy cases, the current major moderators are traditional AI tools and ordinary human reviewers. Even though their accuracy capability has been surpassed by LLMs, platforms still have to be cautious about replacing them with LLMs.

One reason is speed. LLMs generally spend more time in generating decisions because of the larger scale of parameters they have. Ordinary human moderators are generally slower than machines but are quicker than expert moderators, as they spend about 10 to 30 seconds on average in moderating one piece of content (\cite{Gonganeetal2022}, 32; \cite{Gosztonyi2023}, 116). LLM moderation is quicker than human moderation, but takes much more time than its AI predecessors. It is reported that moderating one piece of content could take LLM a few seconds (\cite{Gületal2024}, 9). High latency makes it unlikely for LLMs to fully replace the role of traditional ML models in moderating easy cases, where speed is a key parameter. To be sure, LLMs can supplement traditional ML in some cases, and technical improvement may increase LLMs' speed in the future. But speed is a concern that should not be ignored, especially in real-time moderation tasks.

The second concern is cost. LLM moderation is a costly practice (\cite{Boicel2024}; \cite{Kumaretal2024}, 10). On the one hand, many LLMs, especially their API, are proprietary and charge fees for their uses. On the other hand, fine-tuning the models for specific scenarios introduces extra costs, including computational resources, preparation of the fine-tuning dataset, and the required expertise (\cite{Anggrainingsihetal2024}, 212). Resource intensiveness poses a substantial obstacle for platforms, especially small ones, to put such technique in wide use. In fact, due to the law of diminishing returns, the small accuracy gain brought by LLMs may come at a price of substantial cost. That means, after a certain point, “the cost of reducing the marginal rate of error would become higher and higher... [and platforms would] ...invest enormous resources for an infinitesimal gain in accuracy” (\cite{Solum2004}, 247). Even though LLMs could avoid the manual annotation of traditional ML models, which is labor-intensive and psychologically harmful (\cite{nr_Songetal2025}), the computational cost of LLMs makes it a poor candidate for moderating the massive number of easy cases.

In addition, current studies on LLM moderation either tested the accuracy of LLMs in detecting certain categories of speech (such as hate speech or misinformation) or prompted the LLMs with a particular rule of content. On both occasions, researchers give the LLMs a rule for a specific category of content, and then LLMs are asked to determine whether a post violates that rule. In reality, however, the rules of content in a platform are very complex, covering many kinds of speech; and the first thing LLMs must do is to determine the appropriate rule that is applicable to a piece of content. In other words, all the content rules (tens of thousands of words, at least for a big platform) should be prompted as input to the LLMs. This may further exacerbate the latency of LLMs in moderation tasks. One study shows that once multiple policies have been prompted to LLMs, their accuracy of classification will decrease, and the costs will increase (\cite{Thomasetal2024}, 9-10).

The above analysis suggests that LLMs' diminishing advantage in accuracy compared to expert moderators makes them suitable for directly moderating hard cases. For easy cases, LLMs do generate accuracy dividends for the moderation system, but advocacy for their application should be met with caution because of the increased latency and cost. Thus, LLMs should replace neither expert review in hard cases nor traditional AI and ordinary human moderators in easy cases.

\subsection{\textit{LLMs’ real potential in content moderation}}

If accuracy is not the major field of LLMs' contribution, then what areas can LLMs play a role in? Recognizing LLMs' limited use in accuracy does not dictate their dim prospect in content moderation. As this part will argue, LLMs' major role in content moderation and platform governance is to assist other moderators in gaining legitimacy for the system. Specifically, LLMs can make significant contributions in the following four aspects.

1) To distinguish easy cases and hard cases. This is crucial since the two categories should be assigned with different resources and strategies. LLMs can help with the task of differentiation, conducting preliminary screening, and leaving complex issues to human experts (\cite{nr_Kumaretal2024}; \cite{nr_Villate-Castilloetal2024}). Researchers found that LLMs exhibit satisfactory performance in pre-filtering content, that is, to remove clearly non-violative content from the queue of moderation, as well as to escalate clearly violative content for human review (\cite{Thomasetal2024}, 11). The primary way to conduct this task is by evaluating LLMs' level of confidence or uncertainty (\cite{nr_Chen_Mueller2024}; \cite{Wangetal2024}, 3). For open-sourced LLMs, uncertainty quantification can be measured by leveraging the probability distribution of each token to estimate the probability of responses; while for black-box LLMs, uncertainty can be measured either by evaluating the consistency of the multiple responses or by directly asking the model to estimate its confidence in a verbal way (\cite{nr_Shorinwaetal2024}, 13-18; \cite{nr_Becker_Soatto2024}).

By measuring the level of LLMs' confidence or uncertainty (\cite{Xiongetal2024}), we can tell that if the model is highly uncertain about a case, that case is probably hard. To facilitate LLMs' capacity on this task, we can fine-tune LLMs with an annotated dataset containing easy and hard cases; the reward signal is the ground truth answers (\cite{nr_Stengel-Eskinetal2024}). In practice, LLMs' confidence levels can be adjusted according to changing contexts, such as the need to address emergencies or crises. For instance, if platform managers want to escalate more cases to the "hard" category for human review, the threshold of hard cases can be lowered by adjusting the level of LLMs' confidence score.

Another workable way for LLMs to screen hard cases is through judging by disagreement. For example, human moderators and LLMs can moderate the same content simultaneously; if the two disagree, then it is likely to be a hard case (\cite{Thomasetal2024}, 4). A threshold can be set to the level of human-LLM disagreement for the purpose of judging whether a case a hard (\cite{nr_Guerdanetal2024}).

These findings corroborate LLMs' technical capacity in distinguishing hard cases from easy ones. Due to cost considerations, however, platforms may not choose to screen all the cases through LLMs. Rather, they can either use LLMs to screen only those cases which have been appealed by users, or those which had already been marked as uncertain by traditional ML models. Using LLMs as a second screener can supplement the first reviewer with more contextual knowledge; and instead of replacing all the work of traditional ML models, using the two tools in a collaborative way is also more financially sustainable. Part 7 of this article proposes a tentative workflow.

2) To provide quality explanations for moderation decisions. As a type of generative AI, LLMs' defining feature is text generation. Such a feature, supported by its extensive training, makes it capable of providing high-quality explanations for moderation decisions, enhancing the transparency of the process (\cite{Nghiem_Daumé2024}, 7-8). According to \cite{nr_Wangetal2023}, "the explanations generated by GPT-3 are fluent, informative, persuasive, and logically sound." \cite{nr_Nirmal2024} reported that moderation rationales generated by LLMs are semantically similar to those generated by humans. (\cite{Francoetal2023}, 5) found that the explanations provided by LLMs, though not identical to human reasoning, look quite convincing to users. The reason why LLMs can imitate human explanations is that they are pre-trained by mostly human-authored texts (\cite{nr_Kunz_Kuhlmann2024}). In some occasions, LLM explanations can even be more comprehensive than those written by human annotators (\cite{Huangetal2023}, 4). To improve the relevance and coherence of explanations, LLMs can be fine-tuned with some examples of high-quality human explanations (\cite{nr_Bonaventuraetal2025}).

In the explanation process, LLMs could utilize their character simulation capability to predict audiences' reactions to the moderated content (\cite{nr_Lietal2024}). Adding potential social impact to the meaning of the content itself would make the explanations much more contextual and persuasive. Such explanations can also be dynamic and interactive, including not only the reasons for violating community rules but also recommendations for modification (\cite{nr_Lietal2024}, 17-18). Such an informative process is dialogic and can even start before the content has been made public (\cite{Barrett_Hendrix2024}). This could serve the goal of educating the users and promoting the quality of the online discourse.

LLMs' strength of explainability can be utilized not only in their own moderation process but also for helping justify the decisions made by other moderators, such as the traditional AI tools or the human reviewers (\cite{nr_Kroegeretal2024}). When researchers provide the LLMs with moderation decisions and relevant content rules, it can generate quality explanations for the decisions (\cite{Kollaetal2024}). As has previously argued, it may not be economically feasible for LLMs to directly moderate easy cases; but it can help AI and human moderators generate contextual explanations for their decisions on hard cases. This could save substantial time for other moderators, enhancing the efficiency of the whole system.

One concern for LLMs' capacity to provide explanations comes from their bias, hallucination, and limited ability to understand highly contextual content. This article acknowledges these limits and concurs that LLMs' explanations cannot be relied on at all occasions. To what extent they can be relied upon may depend on various factors, such as the reliability of the model, the level of the cases, and the human moderator that uses the explanation. Even though LLMs' explanations are not perfect, they can still be convincing and well-reasoned, making users generally satisfied. One promising point is that studies found that even though LLMs may not accurately understand contextual content directly, they can generate good explanations for those cases after being provided with moderation outcomes by humans (\cite{nr_Penagos2024}). This can save much labor for human experts. Future research should explore ways of improving the quality and reliability of LLM explanations. In addition, if LLMs' explanations failed to meet users' expectations, LLMs could strive to provide channels for user participation, enabling users to contest those cases. The LLM-enabled contestation could help human experts review those cases. This is another way LLMs can contribute to the moderation process, as illustrated by the fourth point of this section.

3) To assist human reviewers in getting more information or knowledge. Due to extensive pre-training, LLMs' biggest advantage lies in their “comprehensive grounding knowledge, strong language understanding, and logical reasoning capabilities” (\cite{Maetal2024}, 2). The ground knowledge, which has been fed into the LLMs during pre-training by vast text corpora, enables the model to acquire basic understanding across different platforms, contexts, and communities. Such a huge knowledge base empowers LLMs with the ability to provide various sources of information, distinctive from traditional ML models and human moderators, which only acquire domain-specific knowledge (\cite{Shethetal2024}, \cite{nr_Parketal2024}). Empowered by such comprehensive and diverse knowledge, LLMs can be consulted in the moderation process concerning issues such as the factual background of a case, the cultural context, and the delicate value considerations. These types of information can then help the other moderators make more accurate and judicious decisions.

LLMs can help human moderators by providing additional perspectives that they may have ignored. For example, researchers have found that LLMs could "provide human coders with helpful context on US politics, immigration policy debates, and incivil expressions that may be overlooked" (\cite{nr_Parketal2024}). In one experiment, scholars prompted an LLM (GPT-4) to moderate the speech contained in the \textit{Perincek v. Switzerland} case decided by the European Court of Human Rights. The model responds that the speech should not be removed because "additional context was needed to determine if it was part of hate speech or misinformation campaigns" (\cite{nr_Penagos2024}). \cite{nr_Linetal2024}used two LLM agents to moderate different modules of online memes - the debate between them can be extremely helpful for human moderators to judge whether the multi-module meme is harmful. This highlights LLMs' role as an important reference for human moderators, especially in hard cases where extra caution and prudence are needed.

Apart from directly offering information, LLMs can also assist other moderators in several other aspects. For example, LLMs can monitor the algorithmic moderation system by offering statistical insights such as toxicity score (\cite{Sawane2024}) as well as explanations of those metrics. LLMs can also be used to generate datasets to train the traditional ML models. This technique is referred to as data augmentation (\cite{Dingetal2024}). Deploying such capability of LLMs can produce synthetic data, reducing the burden of collecting and labeling data for training the ML models (\cite{Divakaran_Peddinti2024}, 5). Studies revealed that optimized LLMs can generate synthetic datasets with high quality (\cite{nr_Bodaghietal2024}). These are instances where LLMs could boost the functioning of other moderators with their informational competence.

4) To facilitate user participation in a more interactive way. Putting transparency into the framework of governance, we can conclude that “[e]xplanation is not just about providing accurate information about how AI works, but fundamentally social and situated” (\cite{Kou_Gui2020}, 102:18). Transparency is far more than the one-way disclosure of information: it is also a communicative process between the platform and the users (\cite{Suzoretal2019}, 1539). The conversational capability of LLMs can help solicit user participation and feedback on various occasions, such as the appeal of moderation decisions and the revision of content rules. On the one hand, users should have chances to contest individual moderation decisions, including challenging the outcome, debating with rationales, providing new information, and initiating the appeal process. Instead of passively waiting for the platform's reply, users should be engaged with a participatory design that encourages real-time communications (\cite{nr_Vaccaroetal2021}). On the other hand, user participation could go beyond individual cases and cover issues of platform governance in general. One of the key concerns for content moderation is that such private ordering of public rights lacks democratic legitimacy (\cite{nr_Elkin-Koren_Perel2023}). Incorporating users' voices into the governance process, thus, is extremely important. Public participation serves to democratize the moderation process, reveal the hidden value tradeoffs, and check the platforms' exercise of such tremendous power.

LLMs can make a significant contribution to improving such interactive design. The natural language interface of LLMs makes them more convenient and efficient for conversations than previous ML models (\cite{nr_Singhetal2024}). In its interactive API, LLMs could not only receive follow-up feedback from users regarding moderation decisions but also educate users about the rules and norms of the platform. For example, LLMs have been tested to effectively paraphrase hate speech with non-harmful words (\cite{nr_Chaoetal2024}) and to rewrite inappropriate arguments while preserving their content (\cite{nr_Ziegenbeinetal2024}). Such paraphrasing and rewriting capability can be utilized to correct violating behavior and educate users on how to express sentiment without using hateful or derogatory content (\cite{nr_Gabrieletal2024}).

Users could pose their comments on whether the justification of a decision is powerful enough and whether some other factors should be considered in making the decision (\cite{Molina_Sundar2022}, 3). Such an interactive route enhances users’ control and agency in the process (\cite{Molina_Sundar2022}, 9-10). The dialogic feature of LLMs makes them suitable for conducting this task. In particular, LLMs in the conversation can generate personalized responses that are tailored to users' situations and expectations, further improving the experience of participation (\cite{nr_Heetal2024}). User feedback can then be used to further improve the models (\cite{Moscaetal2023}). Empowering users to have their voices heard would greatly enhance the legitimacy of the moderation practice as well as the governance scheme as a whole.

\section{A Proposed Workflow of LLM in Moderation}

From the above illustration of LLMs' potential in content moderation, two important insights could be gained. First, identifying the suitable role of LLMs in content moderation requires us to put this tool within the whole system of platform governance and compare it with other types of moderators. Each type of moderator has its strengths and limits. The table below briefly illustrates the comparison between them.

\begin{table*}[ht]
\centering

\begin{tabular}{| l  |l  |l  |l  |l  |l |} \hline 
  & \multicolumn{3}{|c|}{Easy Cases} & \multicolumn{2}{|c|}{Hard Cases} \\
\hline
  & \textit{Accuracy} & \textit{Speed} & \textit{Transparency} & \textit{Justification} & \textit{Participation} \\
\hline
Traditional ML&   high&   superior&   /&   low&   low\\
\hline
LLM &   high&   high&   /&   medium&   high\\
\hline
Ordinary Human &   high&   medium&   /&   medium&   high\\
\hline
Experts &   superior&   low&   /&   superior&   medium\\ \hline

\end{tabular}

\caption{Different Moderators under the Legitimacy Framework}
\end{table*}

As transparency of easy cases depends on policy choices of the moderation system (subject to regulatory pressures), this metric is largely irrelevant to which type of moderators has been used. In other words, public disclosure of the moderation process is a determination made by top managers of platforms as well as legal regulators, irrespective of which one is doing the moderation. So, the metric of transparency has been left blank in the table. 

Traditional ML models are very quick and cost-effective; they can also achieve sufficiently high accuracy in easy cases. For this reason, the majority of content on most online platforms has been moderated by these tools. However, the ML models cannot offer quality justifications for hard cases due to their limited ability in contextual understanding. These tools are also not interactive, contributing little to user participation.

Ordinary human moderators (mostly outsourced) have high accuracy performance - higher than ML models since they can understand more contexts and nuances. But their speed is much lower than automated tools like ML models and LLMs. In hard cases, human moderators have strength in communicating with users because they are more sympathetic than machine responses. Their capability in justification is higher than ML models (due to better contextual understanding) but not sufficiently high as they are not adequately trained.

Expert reviewers have the best accuracy performance - in many circumstances, their moderation has been treated as the ground truth for labeling. But their speed is extremely low - that's the price to be paid for the careful consideration and deliberation they take. Experts are trained to make sound value judgments with convincing rationales, so their justification is superior. However, due to cost concerns, experts are too few to provide wide participation chances to users.

Compared to those three types of moderators, LLMs may be more accurate than traditional ML models but take more cost and time. LLMs' accuracy advantage over ordinary human moderators is debatable (\cite{nr_Masudetal2024}, \cite{Wengetal2023}), even though the former is quicker than the latter (\cite{nr_González-Bustamante2024}). Viewed in this light, LLMs' contribution should not lie primarily in easy cases, since it has no competitive advantage over the other moderators. In hard cases, however, LLMs can play a significant role. Its capacity for justifications, though obviously cannot meet that of experts, can supplement the work of ML models and ordinary human moderators by providing knowledge and contextual information. Besides, its interactive and conversational feature can facilitate user participation at scale, assisting human moderators and experts. Comparing LLMs' strengths and weaknesses with those of other moderators is necessary for finding a suitable position for LLM moderation. The results of the comparison echo the discussion of LLMs' role in the previous part.

Second, in order to enhance the legitimacy of platform governance under the framework, one crucial strategy is that different types of moderators should not only fulfill their own strengths but also help and supplement other moderators. In other words, the four types of moderators should collaborate and help each other in the governance scheme. Utilizing LLMs for content moderation is not an isolated effort to excavate the new technique’s potential in its own regard but a holistic enterprise that takes the relational structure of governance into account. To grasp how LLMs could be incorporated into the moderation system and collaborate with other moderators, a tentative workflow is proposed here. The workflow aims to serve as a useful reference for platform managers and developers who plan to utilize LLMs in their moderation scheme.

\begin{figure}
    \centering
    \includegraphics[width=1\linewidth]{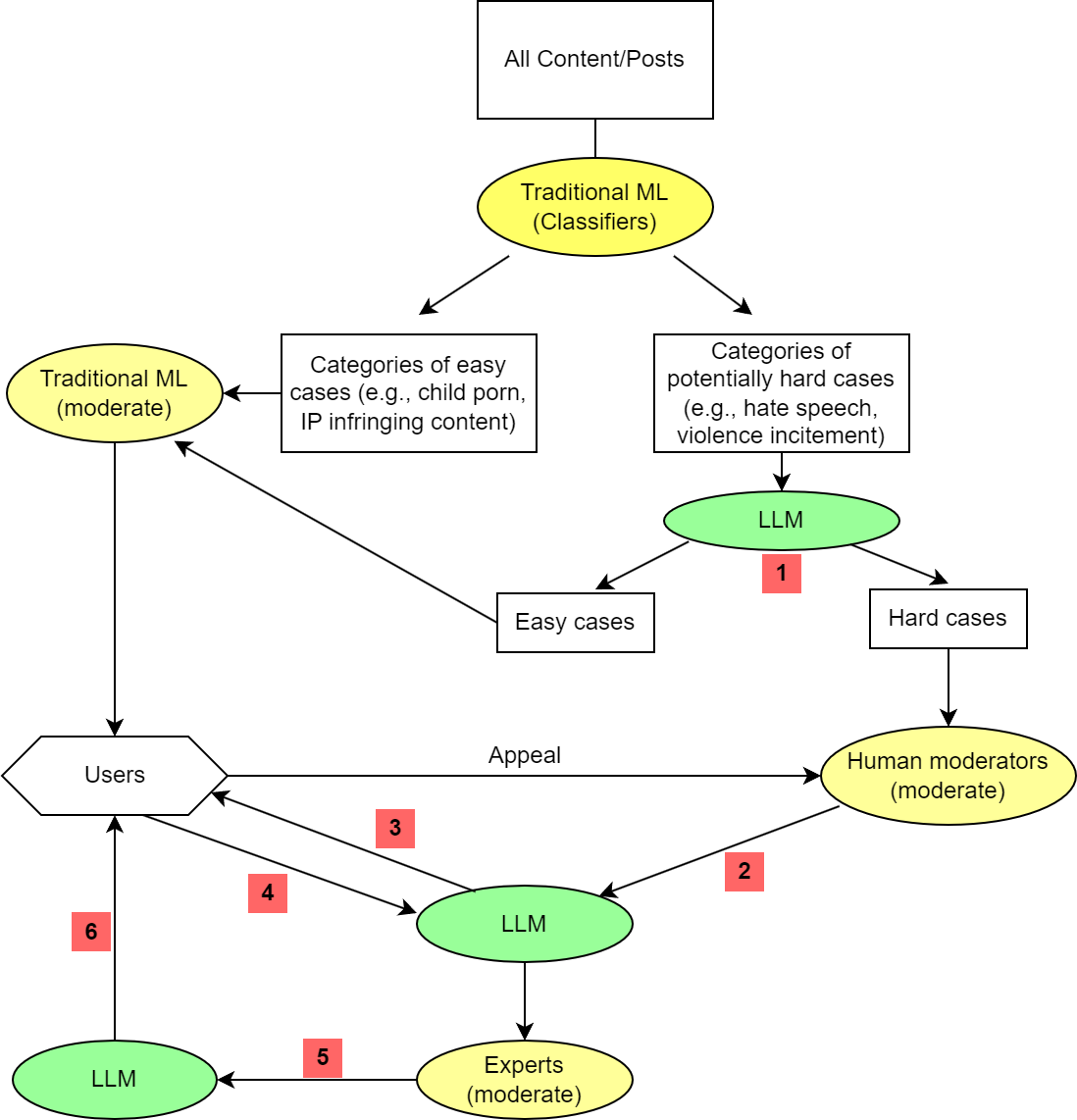}
    \caption{A proposed workflow of incorporating LLM into content moderation. As shown in the figure, LLM can contribute in at least 6 occasions (numbers in red box): 1. distinguishing easy cases and hard cases; 2. providing information to human moderators; 3. helping human moderators generate justification for the moderation; 4. receiving feedback or appeal request from users; 5. assisting experts review by providing information; 6. disclosing results and justifications to users in an interactive way. }
    \label{fig:enter-label}
\end{figure}

The first function played by LLMs in the moderation workflow is distinguishing between easy and hard cases. Here, a typical concern is how to balance costs and benefits. If LLMs bring more costs and latency than traditional ML models and thus are not suitable for replacing the latter in moderating easy cases, how can LLMs be used for prescreening cases and offering explanations? Under my proposed workflow, LLMs do not have to screen all cases, but rather those categories of potentially hard cases, such as hate speech and violence incitement. These types of content often involve vague rules, contextual judgment, and value tradeoffs. By contrast, some categories are less controversial, and the existing ML models have been mature in detecting them, such as child porn and IP-infringing content. Thus, in the first step, traditional ML classifiers can classify content into different categories: ML models can moderate the easy categories and disclose results to users. For the potentially hard cases categories, LLMs can be used to screen them by, e.g., measuring the confidence scores.

After the prescreening, the hard cases and the cases appealed by users will be sent to human moderators. In this step, human moderators could ask LLMs for useful information to help their judgment. After human moderators have reached results, they can use LLMs to generate explanations for disclosure to users. To save cost, one approach is to use an LLM as a teacher to distill knowledge into a smaller student model: empirical study shows that the student model can achieve the equivalent quality of explanation, but it is faster and more cost-efficient (\cite{nr_Piot_Parapar2024}). When users receive the case results and explanations, they can contest the result, appeal the case, or provide feedback through the conversational LLMs.

If users are still unsatisfied with the result, or the case is very important, expert reviewers can get involved in the last step. Similar to the previous step, experts could consult LLMs if they need additional information about the cases. They can also use LLMs to disclose results and justifications to users in an interactive and dialogic way.

This workflow balances LLMs' strengths and limits, as well as considering both costs and benefits. Under this workflow, LLMs' role mainly lies in moderating hard cases rather than increasing accuracy in easy cases (as most existing research focuses on). LLMs' role is also assistive, helping the work of human moderators and experts. Only by collaborating with other moderators can LLMs realize their full potential in the governance of online platforms.

\section{Directions for Future Research}

\subsection{Research on LLM Technology}

Broadening our discourse on LLM moderation from accuracy to legitimacy is the first step. To better facilitate LLMs' contributory role in content moderation and platform governance, endeavors from both technical and legal/policy fields are needed. To some extent, the major contribution of this article is to redirect research of LLM moderation. This section discusses the technical side, leaving law and policy to the next section. The major future work is to empirically test the framework raised by this article (Part 5 and 6) and to improve the tentative workflow (Part 7). Specifically, research on the LLM technology should focus on the four roles that LLMs could play in content moderation. Scholars should further explore different types of LLMs' strengths and weaknesses in the four aspects, in order to find the best available solution for realizing LLMs' potential in the legitimacy framework.

1) Distinguishing between easy and hard cases. The major method for this task is uncertainty measurement of LLMs. Currently, such a technique has plenty of room for improvement. For example, \cite{nr_Leeetal2023} has demonstrated the misalignment between LLM calibration and human disagreement. \cite{nr_Xiongetal2023} discovered that LLMs tend to be overconfident in their responses. Current methods on uncertainty quantification are still far from perfect: token-based entropy for open-source LLMs and response consistency for black-box LLMs are indicators of factual uncertainty, but may not perfectly align with the latter (\cite{nr_Shorinwaetal2024}, 27). Among them, verbalized confidence measurement is an easy-to-use method for it only requires prompting for inquiry, and it is cost-effective; however, its promising application depends heavily on the way LLMs are prompted (\cite{nr_Yangetal2024}). Thus, further exploration of improving the prompting strategy and enhancing the reliability of LLMs' confidence scores is necessary.

2) Providing explanations to users/offering information to other moderators. As these two functions both depend on LLMs' informational capabilities, this section combines them and discusses research needs in these fields.

Human moderators and experts could refer to LLMs to gain more contextual information and knowledge in their moderation process. LLMs can also help other moderators draft justifications for the moderation decisions. Even though LLMs' explainability has been hugely improved in recent years, there still exist various drawbacks. For example, the interpretations generated by LLMs also exhibit non-minimal frequencies of logical error, vagueness, and irrelevance (\cite{nr_Bonaventuraetal2024}). Another feature that compromises the quality of explanations is the sycophancy of LLMs, which tends to please users' views or preferences (\cite{nr_Weietal2024}). Scholars have endeavored to mitigate the sycophantic tendency (\cite{nr_Sharmaetal2023}), but more work needs to be done.

Another significant issue is the bias and hallucination in LLMs' responses. For example, LLMs can reflect and reproduce existing social bias, stereotypes, and discrimination in their explanations (\cite{nr_Cabelloetal2023}). And these biases can have a strong misleading effect on humans (\cite{nr_Wangetal2023}). LLMs have been found to be geographically (\cite{nr_Manvietal2024}) and politically (\cite{nr_Motokietal2024}) biased - such feature is especially troublesome for maintaining the global and inclusive online speech platforms. Hallucinations are the nonsensical or fabricated answers produced by LLMs; they are not fully explainable: On many occasions, developers do not know why they emerge and how to prevent them (\cite{Yaoetal2024}). Bias and hallucinations can mislead people since even false or fabricated responses are highly convincing.

To mitigate this risk, more efforts are needed. First, techniques of detecting, measuring, and mitigating LLMs' bias and hallucination are waiting to be improved. Current endeavors include prompt engineering, supervised fine-tuning, designing multiple LLM agents to debate their claims, and introducing human-in-the-loop (\cite{nr_YueZhangetal2023}, \cite{nr_JunyiLietal2024}). Second, human moderators and experts should be trained in meticulously using LLMs to avoid being misled by the models; they should have extra sources of fact-checking to verify the factual claims generated by LLMs. Third, there must be policies regulating the potential risks of bias in the use of LLMs (\cite{nr_Dasetal2024}), e.g., by mandating human oversight and public audit. In the future, research should be devoted to enhancing the relevance, persuasiveness, and trustworthiness of LLMs' explanations (\cite{nr_Fergusonetal2024}), as well as to explore better ways of incorporating such justificatory approaches into the platform governance system, supplementing the work of human moderators. Investigating new ways of reducing LLMs' bias and hallucinations, especially in hard cases that are highly contextual, is urgently needed.

4) Facilitating user participation. LLMs' role in this aspect derives from their conversational nature. The key direction for future research is to design more user-centric applications and interfaces and to make them compliant with legal regulations on user rights. Routes, channels, and affordances of user engagement should be built into the API in an accessible manner to facilitate participation and contestation by design (\cite{nr_Vaccaroetal2021}). Besides, many regulations have prescribed users' participatory rights, such as the right to rectify and the right to appeal. Future developers should pay attention to the relevant regulatory requirements in designing LLMs. 

\subsection{Research on Law/Policy}

On the legal/policy side, at least three issues need further research. First, legal regulations must strike a balance between the needs of moderation \textit{by} LLMs and moderation \textit{of} LLMs. To mitigate the harmful content generated by LLM, moderation of their output is needed. One effective approach is to conduct curation during LLMs' pre-training process (\cite{Liuetal2024}); another approach is to add safety filters in LLMs' API (\cite{Lietal2024}, 24). The curation and filtering are used to prevent malicious content from being fed into the model. But they would also compromise the model’s capability of conducting moderation tasks, because lack of exposure to harmful content makes LLMs less effective in recognizing and classifying these types of content. Researchers also found that LLMs could refuse to engage in harmful content in classification tasks (\cite{nr_Dönmezetal2024}). That means there is a tension between promoting safety and preserving the integrity of the training data (\cite{Razaetal2024}). If laws impose strict requirements upon the health and safety of LLMs, then their ability to moderate will likely decline. Lawmakers and policymakers must reach a delicate balance between the two competing goals.

Second, even though LLMs exhibit impressive performance in explaining decisions, the explanations generated by them are not always reliable. Sometimes, LLMs may be uncertain about their reasoning, expressing confusion like a human (\cite{Huangetal2023}, 2). And like the previous technical tools, LLMs also contain the risk of cloaking errors and biases with techno-objectivity. When LLMs provide seemingly convincing but actually misleading explanations for their decisions (\cite{Huangetal2023}, 3), such explanations cannot be counted as valid justification. To address this issue, regulators could consider either mandating the right of users to contest LLM explanations or requiring more transparency on the development of the model to facilitate easier identification of false and misleading explanations.

Third, incorporating LLMs into the system of platform governance demands corresponding regulatory and public oversight. Remember that LLM is itself a tool of power and a site for control. When it is used for governing online spaces, it becomes more than urgent to limit and oversee such formidable power. LLMs contain bias, generate hallucinations, and are vulnerable to manipulations (\cite{Weidingeretal2021}). How can we ensure that the LLM moderator, as a component of the platforms’ governance scheme, is accountable to the public ends and serves legitimate goals? This is a question that regulators should not wait to tackle. Liability law is one tool, for it can provide strong incentives. The specific rules of LLM liability should take into account the specific application scenarios and their stakes on legitimacy. The aim is to incentivize LLMs to be used in a responsible way that enhances, rather than threatens, the legitimacy of platform governance.

\section{Conclusion}

Using LLMs for content moderation is an exciting field to work on. Realizing LLMs' full potential in moderation tasks depends upon locating this technology within the governance structure of online platforms and communities. The currently dominant discourse on LLM moderation focuses on accuracy – how to deploy the tool to increase the ratio of correct decisions. However, this article argues that accuracy is not the best area of contribution by LLMs. LLMs' accuracy advantage as compared to traditional AI and ordinary human moderators has been largely offset by its weakness in cost and latency. Rather, LLMs can make meaningful contributions in other aspects, such as distinguishing hard cases from easy ones and providing interactive channels for user participation. Moving from accuracy to legitimacy, we can get a clearer picture of LLMs' role in moderation and governance.

The critical analysis offered by this article affirms the necessity of combining technical explorations with normative inquiries from the socio-legal perspective. If the objective of LLM moderation is to assist online platforms to better govern their communicative space, then the research effort should not be fixated upon accuracy only. The literature from law and social sciences, such as the studies on platform governance, the division between easy and hard cases, and the conceptualization of legitimacy, supplies valuable insights to the research field of LLM moderation. Such an interdisciplinary approach is indispensable for future studies of technologies and their impact on human society.

\section{Statements and Declarations}
The authors have no competing interests to declare that are relevant to the content of this article.

The work described in this paper was fully supported by a grant from City University of Hong Kong (Project No. 7200786).

\printbibliography

@article{Tyleretal2015,
    author = "Tom Tyler and Phillip Atiba Goff and Robert J. MacCoun",
    title = "The Impact of Psychological Science on Policing in the United States: Procedural Justice, Legitimacy, and Effective Law Enforcement",
    journal = "Psychological Science in the Public Interest",
    Volume = "16",
    number = "3",
    pages = "75-109",
    year = "2015"
}

@article{Jhaveretal2019,
    author = "Shagun Jhaver and Amy Bruckman and Eric Gilbert",
    title = "Does Transparency in Moderation Really Matter?: User Behavior After Content Removal Explanations on Reddit",
    journal = "Proc. ACM Hum.-Comput. Interact.",
    Volume = "3",
    number = "CSCW",
    pages = "150",
    year = "2019"
}

@misc{Masnick2019,
    author    = "Mike Masnick",
    title     = "Masnick's Impossibility Theorem: Content Moderation At Scale Is Impossible To Do Well",
    url       = "https://www.techdirt.com/2019/11/20/masnicks-impossibility-theorem-content-moderation-scale-is-impossible-to-do-well/",
    addendum = "(accessed: Aug 13, 2024)",
}

@article{Suzoretal2019,
    author = "Nicolas Suzor and Sarah West and Andrew Quodling and Jillian York",
    title = "What Do We Mean When We Talk About Transparency? Toward Meaningful Transparency in Commercial Content Moderation",
    journal = "International Journal of Communication",
    Volume = "13",
    pages = "1526-1543",
    year = "2019"
}

@article{Kou_Gui2020,
    author = "Yubo Kou and Xinning Gui",
    title = "Mediating Community AI Interaction through Situated Explanation: The Case of AI Led Moderation",
    journal = "Proceedings of the ACM on Human-Computer Interaction",
    volume = "4",
    number = "CSCW2",
    page = "102",
    year = "2020",
    DOI = "https://doi.org/10.1145/3415173"
}

@article{OConnor_Joffe2020,
    author = "Cliodhna O’Connor and Helene Joffe",
    title = "Intercoder Reliability in Qualitative Research: Debates and Practical Guidelines",
    journal = "International Journal of Qualitative Methods",
    volume = "19",
    pages = "1-13",
    year = "2020",
    DOI = "10.1177/1609406919899220"
}

@online{Dentonetal2021,
    author    = "Remi Denton and Mark Díaz and Ian Kivlichan and Vinodkumar Prabhakaran and Rachel Rosen",
    title     = "Whose Ground Truth? Accounting for Individual and Collective Identities Underlying Dataset Annotation",
    url       = "https://arxiv.org/abs/2112.04554",
    year = "2021"
}

@misc{Weidingeretal2021,
    author    = "Laura Weidinger et. al.",
    title     = "Ethical and social risks of harm from Language Models",
    url       = "https://arxiv.org/abs/2112.04359",
    year = "2021"
}

@article{Molina_Sundar2022,
    author = "Maria D. Molina and S. Shyam Sundar",
    title = "When AI moderates online content: effects of human collaboration and interactive transparency on user trust",
    journal = "Journal of Computer-Mediated Communication",
    volume = "27",
    number = "4",
    pages = "1-12",
    year = "2022",
    DOI = "https://doi.org/10.1093/jcmc/zmac010"
}

@misc{DAlonzo2023,
    author    = "Marissa D'Alonzo",
    title     = "What is AI anyway, and why should I care?",
    url       = "https://sciencefortheunscientific.substack.com/p/what-is-ai-anyways-and-why-should",
    year = "2023",
    addendum = "(accessed: Aug 13, 2024)"
}

@article{Francoetal2023,
    author = "Mirko Franco and Ombretta Gaggi and Claudio E. Palazzi",
    title = "Analyzing the Use of Large Language Models for Content Moderation with ChatGPT Examples",
    journal = "Proceedings of the 3rd International Workshop on Open Challenges in Online Social Networks",
    pages = "1-8",
    year = "2023",
    DOI = "10.1145/3599696.3612895"
}

@online{Gilardietal2023,
    author    = "Fabrizio Gilardi and Meysam Alizadeh and Maël Kubli",
    title     = "ChatGPT Outperforms Crowd-Workers for Text-Annotation Tasks",
    url       = "https://arxiv.org/abs/2303.15056",
    year = "2023"
}

@inbook{Gosztonyi2023,
   author = "Gergely Gosztonyi",
   title = "Censorship from Plato to Social Media. Law, Governance and Technology Series",
   publisher = "Springer",
   year = "2023",
   chapter = "8. Human and Technical Aspect of Content Management",
   page = "111"
}

@online{Huangetal2023,
    author    = "Fan Huang and Haewoon Kwak and Jisun An",
    title     = "Is ChatGPT better than Human Annotators? Potential and Limitations of ChatGPT in Explaining Implicit Hate Speech",
    url       = "https://arxiv.org/abs/2302.07736",
    year = "2023"
}

@online{Menon2023,
    author    = "Pradeep Menon",
    title     = "Introduction to Large Language Models and the Transformer Architecture",
    url       = "https://rpradeepmenon.medium.com/introduction-to-large-language-models-and-the-transformer-architecture-534408ed7e61",
    year = "2023",
    addendum = "(accessed: Aug 13, 2024)"
}

@article{Moketal2023,
    author = "Lillio Mok and Sasha Manda and Ashton Anderson",
    title = "People Perceive Algorithmic Assessments as Less Fair and Trustworthy Than Identical Human Assessments",
    journal = "Proc. ACM Hum. Comput. Interact.",
    volume = "7",
    page = "309",
    year = "2023",
    DOI = "10.1145/3610100"
}

@online{Moscaetal2023,
    author    = "Edoardo Mosca et. al.",
    title     = "IFAN: An Explainability-Focused Interaction Framework for Humans and NLP Models",
    url       = "https://arxiv.org/abs/2303.03124",
    year = "2023"
}

@article{Mullicketal2023,
    author = "Sankha Subhra Mullick et. al.",
    title = "Content Moderation for Evolving Policies using Binary Question Answering",
    journal = "Proceedings of the 61st Annual Meeting of the Association for Computational Linguistics",
    volume = "5",
    pages = "561-573",
    year = "2023",
    DOI = "10.18653/v1/2023.acl-industry.54"
}

@online{Ottosson2023,
    author    = "Dan Ottosson",
    title     = "Cyberbullying Detection on social platforms using Large Language Models",
    url       = "https://www.diva-portal.org/smash/get/diva2:1786271/FULLTEXT01.pdf",
    year = "2023"
}

@online{Royetal2023,
    author    = "Sarthak Roy and Ashish Harshavardhan and Animesh Mukherjee and Punyajoy Saha",
    title     = "Probing LLMs for hate speech detection: strengths and vulnerabilities",
    url       = "https://arxiv.org/abs/2310.12860",
    year = "2023"
}

@online{Vaswanietal2023,
    author    = "Ashish Vaswani et. al.",
    title     = "Attention Is All You Need",
    url       = "https://arxiv.org/abs/1706.03762",
    year = "2023"
}

@online{Wangetal2023,
    author    = "Gordon Wang and Yanwei Cui and Melanie Li",
    title     = "Build a generative AI-based content moderationsolution on Amazon SageMaker JumpStart",
    url       = "https://aws.amazon.com/cn/blogs/machine-learning/build-a-generative-ai-based-content-moderation-solution-on-amazon-sagemaker-jumpstart/",
    year = "2023",
    addendum = "(accessed: Aug 13, 2024)"
}

@online{Wengetal2023,
    author    = "Lilian Weng and Vik Goel and Andrea Vallone",
    title     = "Using GPT-4 for content moderation",
    url       = "https://openai.com/index/using-gpt-4-for-content-moderation/",
    year = "2023"
}

@article{Anggrainingsihetal2024,
    author = "Rini Anggrainingsih and Ghulam Mubashar Hassan and Amitava Datta",
    title = "Transformer -- based models for combating rumours on microblogging platforms: a review",
    journal = "Artificial Intelligence Review",
    volume = "57",
    page = "212",
    year = "2024",
    DOI = "10.1007/s10462-024-10837-9"
}

@online{Barrett_Hendrix2024,
    author    = "Paul M. Barrett and Justin Hendrix",
    title     = "Is Generative AI the Answer for theFailures of Content Moderation?",
    url       = "https://www.justsecurity.org/94118/is-generative-ai-the-answer-for-the-failures-of-content-moderation/",
    year = "2024",
    addendum = "(accessed: Aug 13, 2024)"
}

@article{Bhattacharyaetal2024,
    author = "Pronaya Bhattacharya et. al.",
    title = "Demystifying ChatGPT: An In‑depth Survey of OpenAI’s Robust Large Language Models",
    journal = "Archives of Computational Methods in Engineering",
    year = "2024",
    DOI = "10.1007/s11831-024-10115-5"
}

@online{Boicel2024,
    author    = "Alyssa Boicel",
    title     = "Using LLMs to Moderate Content: Are TheyReady for Commercial Use?",
    url       = "https://www.techpolicy.press/using-llms-to-moderate-content-are-they-ready-for-commercial-use/",
    year = "2024",
    addendum = "(accessed: Aug 13, 2024)"
}

@article{Choi_Ferrara2024,
    author = "Eun Cheol Choi and Emilio Ferrara",
    title = "Automated Claim Matching with Large Language Models: Empowering Fact-Checkers in the Fight Against Misinformation",
    journal = "WWW ’24 Companion",
    year = "2024",
    DOI = "10.1145/3589335.3651910"
}

@online{Dingetal2024,
    author    = "Bosheng Ding et. al.",
    title     = "Data Augmentation using Large Language Models: Data Perspectives, Learning Paradigms and Challenges",
    url       = "https://arxiv.org/abs/2403.02990",
    year = "2024"
}

@online{Divakaran_Peddinti2024,
    author    = "Dinil Mon Divakaran and Sai Teja Peddinti",
    title     = "LLMs for Cyber Security: New Opportunities",
    url       = "https://arxiv.org/abs/2404.11338",
    year = "2024"
}

@online{Gületal2024,
    author    = "Ilker Gül and Rémi Lebret and Karl Aberer",
    title     = "Stance Detection on Social Media with Fine-Tuned Large Language Models",
    url       = "https://arxiv.org/abs/2404.12171",
    year = "2024"
}

@article{Katsarosetal2024,
    author = "Matthew Katsaros and Jisu Kim and Tom Tyler",
    title = "Online Content Moderation: Does Justice Need a Human Face?",
    journal = "International Journal of Human–Computer Interaction",
    volume = "40",
    pages = "66-77",
    year = "2024",
    DOI = "10.1080/10447318.2023.2210879"
}

@online{Kokaetal2024,
    author    = "Sahas Koka and Anthony Vuong and Anish Kataria",
    title     = "Evaluating the Efficacy of Large Language Models in Detecting Fake News: A Comparative Analysis",
    url       = "https://arxiv.org/abs/2406.06584",
    year = "2024"
}

@article{Kollaetal2024,
    author = "Mahi Kolla and Siddharth Salunkhe and Eshwar Chandrasekharan and Koustuv Saha",
    title = "LLM-Mod: Can Large Language Models Assist Content Moderation?",
    journal = "CHI EA '24: Extended Abstracts of the CHI Conference on Human Factors in Computing Systems",
    pages = "1-8",
    year = "2024",
    DOI = "10.1145/3613905.3650828"
}

@online{Kumaretal2024,
    author    = "Deepak Kumar and Yousef AbuHashem and Zakir Durumeric",
    title     = "Watch Your Language: Investigating Content Moderation with Large Language Models",
    url       = "https://arxiv.org/abs/2309.14517",
    year = "2024"
}

@article{Lietal2024,
    author = "Lingyao Li and Lizhou Fan and Shubham Atreja and Libby Hemphill",
    title = "“HOT” ChatGPT:The Promise of ChatGPT in Detecting and Discriminating Hateful, Offensive, and Toxic Comments on Social Media",
    journal = "ACM Transactions on the Web",
    volume = "18",
    pages = "1-36",
    year = "2024",
    DOI = "10.1145/3643829"
}

@online{Liuetal2024,
    author    = "Xiaoqun Liu and Jiacheng Liang and Muchao Ye and Zhaohan Xi",
    title     = "Robustifying Safety-Aligned Large Language Models through Clean Data Curation",
    url       = "https://arxiv.org/abs/2405.19358",
    year = "2024"
}

@online{Maetal2024,
    author    = "Huan Ma and Changqing Zhang and Huazhu Fu and Peilin Zhao and Bingzhe Wu",
    title     = "Adapting Large Language Models for Content Moderation: Pitfalls in Data Engineering and Supervised Fine-tuning",
    url       = "https://arxiv.org/abs/2310.03400",
    year = "2024"
}

@online{Nghiem_Daumé2024,
    author    = "Huy Nghiem and Hal Daumé III",
    title     = "HateCOT: An Explanation-Enhanced Dataset for Generalizable Offensive Speech Detection via Large Language Models",
    url       = "https://arxiv.org/abs/2403.11456v2",
    year = "2024"
}

@article{Quelle_Bovet2024,
    author = "Dorian Quelle and Alexandre Bovet",
    title = "The perils and promises of fact-checking with large language models",
    journal = "Front. Artif. Intell.",
    volume = "7",
    page = "1341697",
    year = "2024",
    DOI = "10.3389/frai.2024.1341697"
}

@article{Rawatetal2024,
    author = "Anchal Rawat and Santosh Kumar and Surender Singh Samant",
    title = "Hate speech detection in social media: Techniques, recent trends, and future challenges",
    journal = "WIREs Computational Statistics",
    volume = "16",
    page = "1648",
    year = "2024",
    DOI = "10.1002/wics.1648"
}

@online{Razaetal2024,
    author    = "Shaina Raza and Ananya Raval and Veronica Chatrath",
    title     = "MBIAS: Mitigating Bias in Large Language Models While Retaining Context",
    url       = "https://arxiv.org/abs/2405.11290",
    year = "2024"
}

@online{Sawane2024,
    author    = "Swarnim Sawane",
    title     = "Content Moderation using AI",
    url       = "https://www.cloudraft.io/blog/content-moderation-using-llamaindex-and-llm",
    year = "2024",
    addendum = "(accessed: Aug 13, 2024)"
}

@article{Shethetal2024,
    author = "Paras Sheth and Raha Moraffah and Tharindu S. Kumarage and Aman Chadha and Huan Liu",
    title = "Causality Guided Disentanglement for Cross-Platform Hate Speech Detection",
    journal = "WSDM '24: Proceedings of the 17th ACM International Conference on Web Search and Data Mining",
    pages = "626-635",
    year = "2024",
    DOI = "10.1145/3616855.3635771"
}

@article{Surden2024,
    author = "Harry Surden",
    title = "ChatGPT, AI Large Language Models, And Law",
    journal = "Fordham Law Review",
    volume = "92",
    page = "1941",
    year = "2024",
}

@online{Thomasetal2024,
    author    = "Kurt Thomas et. al.",
    title     = "Supporting Human Raters with the Detection of Harmful Content using Large Language Models",
    url       = "https://arxiv.org/abs/2406.12800",
    year = "2024",
}

@online{Vishwamitraetal2024,
    author    = "Nishant Vishwamitra et. al.",
    title     = "Moderating New Waves of Online Hate with Chain-of-Thought Reasoning in Large Language Models",
    url       = "https://arxiv.org/abs/2312.15099",
    year = "2024",
}

@article{Wangetal2024,
    author = "Xinru Wang and Hannah Kim and Sajjadur Rahman and Kushan Mitra and Zhengjie Miao",
    title = "Human-LLM Collaborative Annotation Through Effective Verification of LLM Labels",
    journal = "CHI '24: Proceedings of the CHI Conference on Human Factors in Computing Systems",
    pages = "1-21",
    year = "2024",
    DOI = "10.1145/3613904.3641960"
}

@online{Willner_Chakrabarti2024,
    author    = "Dave Willner and Samidh Chakrabarti",
    title     = "Using LLMs for Policy-Driven Content Classification",
    url       = "https://www.techpolicy.press/using-llms-for-policy-driven-content-classification/",
    year = "2024",
    addendum = "(accessed: Aug 13, 2024)"
}

@article{Xiongetal2024,
    author = "Miao Xiong et. al.",
    title = "Can LLMs Express Their Uncertainty? An Empirical Evaluation of Confidence Elicitation in LLMs",
    journal = "The 12th International Conference on Learning Representations",
    year = "2024",
    url = "https://openreview.net/forum?id=gjeQKFxFpZ"
}

@online{Yaoetal2024,
    author    = "Jia-Yu Yao and Kun-Peng Ning and Zhen-Hui Liu and Mu-Nan Ning and Yu-Yang Liu and Li Yuan",
    title     = "LLM Lies: Hallucinations are not Bugs, but Features as Adversarial Examples",
    url       = "https://arxiv.org/abs/2310.01469",
    year = "2024",
}

@online{Akre2023,
    author    = "Karin Akre",
    title     = "F-score",
    url       = "https://www.britannica.com/science/F-score",
    year = "2024",
    addendum = "(accessed: Aug 13, 2024)"
}

@online{Fried2023,
    author    = "Ina Fried",
    title     = "OpenAI touts GPT-4 for content moderation",
    url       = "https://www.axios.com/2023/08/15/openai-touts-gpt-4-for-content-moderation",
    year = "2024",
    addendum = "(accessed: Aug 13, 2024)"
}

@book{Waldron1999,
    author    = "Jeremy Waldron",
    title     = "Law and Disagreement",
    year      = "1999",
    publisher = "Oxford University Press",
}

@book{Bradford2020,
    author    = "Anu Bradford",
    title     = "The Brussels Effect: How the European Union Rules the World",
    year      = "2020",
    publisher = "Oxford University Press",
}

@book{Gillespie2018,
    author    = "Tarleton Gillespie",
    title     = "Custodians of the Internet: Platforms, Content Moderation, and the Hidden Decisions That Shape Social Media",
    year      = "2018",
    publisher = "Yale University Press",
}

@article{Gillespie2020,
    author = "Tarleton Gillespie",
    title = "Content moderation, AI, and the question of scale",
    journal = "Big Data \& Society",
    volume = "7",
    year = "2020",
    DOI = "10.1177/2053951720943234"
}

@online{Llansóetal2020,
    author    = "Emma Llansó and Joris van Hoboken and Paddy Leerssen and Jaron Harambam",
    title     = "Artificial Intelligence, Content Moderation, and Freedom of Expression",
    url       = "https://www.ivir.nl/publicaties/download/AI-Llanso-Van-Hoboken-Feb-2020.pdf",
    year = "2020",
    addendum = "(accessed: Aug 13, 2024)"
}

@online{Sartor_Loreggia2020,
    author    = "Giovanni Sartor and Andrea Loreggia",
    title     = "The impact of algorithms for online content filtering or moderation",
    url       = "https://www.europarl.europa.eu/thinktank/en/document/IPOL_STU(2020)657101",
    year = "2020",
    addendum = "(accessed: Aug 13, 2024)"
}

@article{Gonganeetal2022,
    author = "Vaishali U. Gongane and Mousami V. Munot1 and Alwin D. Anuse",
    title = "Detection and moderation of detrimental content on social media platforms: current status and future directions",
    journal = "Social Network Analysis and Mining",
    volume = "12",
    page = "12",
    year = "2022",
    DOI = "10.1007/s13278-022-00951-3"
}

@online{Meta2022b,
    author    = "Meta",
    title     = "How review teams work",
    url       = "https://transparency.meta.com/enforcement/detecting-violations/how-review-teams-work/",
    year = "2022",
    addendum = "(accessed: Aug 13, 2024)"
}

@online{Meta2022c,
    author    = "Meta",
    title     = "How Meta prioritizes content for review",
    url       = "https://transparency.meta.com/policies/improving/prioritizing-content-review/",
    year = "2022",
    addendum = "(accessed: Aug 13, 2024)"
}

@article{Marsoofetal2023,
    author = "Althaf Marsoof and Andrés Luco and Harry Tan and Shafiq Joty",
    title = "Content-filtering AI systems–limitations, challenges and regulatory approaches",
    journal = "Information \& Communications Technology Law",
    volume = "32",
    pages = "64-101",
    year = "2023",
    DOI = "10.1080/13600834.2022.2078395"
}

@online{Rasoetal2018,
    author    = "Filippo Raso and Hannah Hilligoss and Vivek Krishnamurthy and Christopher Bavitz and Levin Kim",
    title     = "Artificial Intelligence \& Human Rights: Opportunities \& Risks",
    url       = "https://dash.harvard.edu/handle/1/38021439",
    year = "2018",
    addendum = "(accessed: Aug 13, 2024)"
}

@article{Hart1958,
    author = "H. L. A. Hart",
    title = "Positivism and the Separation of Law and Morals",
    journal = "Harvard Law Review",
    volume = "71",
    pages = "593-629",
    year = "1958",
}

@article{Dworkin1975,
    author = "Ronald Dworkin",
    title = "Hard Cases",
    journal = "Harvard Law Review",
    volume = "88",
    pages = "1057-1109",
    year = "1975",
}

@article{Soper1977,
    author = "E. Philip Soper",
    title = "Legal Theory and the Obligation of a Judge: The Hart/Dworkin Dispute",
    journal = "Michigan Law Review",
    volume = "75",
    pages = "473-519",
    year = "1977",
}

@article{Hutchinson_Wakefield1982,
    author = "Alan Hutchinson and John Wakefield",
    title = "A Hard Look at Hard Cases: The Nightmare of a Noble Dreamer",
    journal = "Oxford Journal of Legal Studies",
    volume = "2",
    pages = "86-110",
    year = "1982",
}

@article{Schauer1985,
    author = "Frederick Schauer",
    title = "Easy Cases",
    journal = "Southern California Law Review",
    volume = "58",
    pages = "399-440",
    year = "1985",
}

@article{Hutchinson1989,
    author = "Allan C. Hutchinson",
    title = "Democracy and Determinacy: An Essay on Legal Interpretation",
    journal = "University of Miami Law Review",
    volume = "43",
    pages = "541-576",
    year = "1989",
}

@online{Kleinman2016,
    author    = "Zoe Kleinman",
    title     = "Fury over Facebook 'Napalm girl' censorship",
    url       = "https://www.bbc.com/news/technology-37318031",
    year = "2016",
    addendum = "(accessed: Aug 13, 2024)"
}

@article{Crowe2019,
    author = "Jonathan Crowe",
    title = "Not-So-Easy Cases",
    journal = "Statute Law Review",
    volume = "40",
    pages = "75-86",
    year = "2019",
    DOI = "10.1093/slr/hmy027"
}

@online{Dwoskin2020,
    author    = "Elizabeth Dwoskin",
    title     = "Mark Zuckerberg’s reversal on Holocaustdenial is a 180",
    url       = "https://www.washingtonpost.com/technology/2020/10/12/zuckerberg-holocaust-denial-facebook/",
    year = "2020",
    addendum = "(accessed: Aug 13, 2024)"
}

@article{Fischman2021,
    author = "Joshua B. Fischman",
    title = "How Many Cases Are Easy?",
    journal = "Journal of Legal Analysis",
    volume = "13",
    pages = "595-656",
    year = "2021",
    DOI = "10.1093/jla/laaa010"
}

@online{Demopoulos2023,
    author    = "Alaina Demopoulos",
    title     = "Free the nipple: Facebook and Instagramtold to overhaul ban on bare breasts",
    url       = "https://www.theguardian.com/technology/2023/jan/17/free-the-nipple-meta-facebook-instagram",
    year = "2023",
    addendum = "(accessed: Aug 13, 2024)"
}

@article{Watson2023,
    author = "Bill Watson",
    title = "Explaining legal agreement",
    journal = "Jurisprudence",
    volume = "14",
    pages = "221-253",
    year = "2023",
    DOI = "10.1080/20403313.2023.2165789"
}

@article{Mondak1994,
    author = "Jeffery Mondak",
    title = "Policy Legitimacy and the Supreme Court: The Sources and Contexts of Legitimation",
    journal = "Political Research Quarterly",
    volume = "47",
    pages = "675-692",
    year = "1994",
}

@article{Freeman2000,
    author = "Jody Freeman",
    title = "The Private Role in Public Governance",
    journal = "New York University Law Review",
    volume = "75",
    pages = "543-675",
    year = "2000",
}

@article{Solum2004,
    author = "Lawrence B. Solum",
    title = "Procedural Justice",
    journal = "Southern California Law Review",
    volume = "78",
    pages = "181-321",
    year = "2004",
}

@article{Brennan-Marquez2017,
    author = "Kiel Brennan-Marquez",
    title = "'Plausible Cause': Explanatory Standards in the Age of Powerful Machines",
    journal = "Vanderbilt Law Review",
    volume = "70",
    pages = "1249-1301",
    year = "2017",
}

@article{Klonick2018,
    author = "Kate Klonick",
    title = "The New Governors: The People, Rules, and Processes Governing Online Speech",
    journal = "Harvard Law Review",
    volume = "131",
    pages = "1598-1670",
    year = "2018",
}

@article{Suzor2018,
    author = "Nicolas Suzor",
    title = "Digital Constitutionalism: Using the Rule of Law to Evaluate the Legitimacy of Governance by Platforms",
    journal = "Social Media + Society",
    volume = "4",
    year = "2018",
    DOI = "10.1177/2056305118787812"
}

@article{Suzoretal2018,
    author = "Nicolas Suzor and Tess Van Geelen and Sarah Myers West",
    title = "Evaluating the legitimacy of platform governance: A review of research and a shared research agenda",
    journal = "the International Communication Gazette",
    volume = "80",
    pages = "385-400",
    year = "2018",
    DOI = "10.1177/1748048518757142"
}

@online{Ananny2019,
    author    = "Mike Ananny",
    title     = "Probably Speech, Maybe Free: Toward a Probabilistic Understanding of Online Expression and Platform Governance",
    url       = "https://knightcolumbia.org/content/probably-speech-maybe-free-toward-a-probabilistic-understanding-of-online-expression-and-platform-governance",
    year = "2019",
    addendum = "(accessed: Aug 13, 2024)"
}

@article{Bagley2019,
    author = "Nicholas Bagley",
    title = "The Procedure Fetish",
    journal = "Michigan Law Review",
    volume = "118",
    pages = "345-402",
    year = "2019",
}

@article{Bloch-Wehba2019,
    author = "Hannah Bloch-Wehba",
    title = "Global Platform Governance: Private Power in the Shadow of the State",
    journal = "SMU Law Review",
    volume = "72",
    pages = "27-80",
    year = "2019",
}

@online{Bradfordetal2019,
    author    = "Ben Bradford et. al.",
    title     = "Report Of The Facebook Data Transparency Advisory Group",
    url       = "https://www.justicehappenshere.yale.edu/reports/report-of-the-facebook-data-transparency-advisory-group",
    year = "2019",
    addendum = "(accessed: Aug 13, 2024)"
}

@article{Kaminski2019,
    author = "Margot E. Kaminski",
    title = "Binary Governance: Lessons from the GDPR’s Approach to Algorithmic Accountability",
    journal = "Southern California Law Review",
    volume = "92",
    pages = "1529-1616",
    year = "2019",
}

@online{FacebookOversightBoard2020,
    author    = "Facebook Oversight Board",
    title     = "OVERSIGHT BOARD OVERTURNS FACEBOOK DECISION IN NAZI QUOTE CASE",
    url       = "https://www.oversightboard.com/news/141077647749726-oversight-board-overturns-facebook-decision-case-2020-005-fb-ua/",
    year = "2020",
    addendum = "(accessed: Aug 13, 2024)"
}

@article{Douek2021,
    author = "Evelyn Douek",
    title = "Governing Online Speech: From 'Posts-As-Trumps' to Proportionality and Probability",
    journal = "Columbia Law Review",
    volume = "121",
    page = "759-833",
    year = "2021",
}

@article{Douek2022,
    author = "Evelyn Douek",
    title = "Content Moderation as Systems Thinking",
    journal = "Harvard Law Review",
    volume = "136",
    page = "526-607",
    year = "2022",
}

@online{MetaOversightBoard2022,
    author    = "Meta Oversight Board",
    title     = "Donation of Pharmaceutical Drugs to Sri Lanka",
    url       = "https://transparency.meta.com/oversight/oversight-board-cases/donations-of-pharmaceutical-drugs-sri-lanka",
    year = "2020",
    addendum = "(accessed: Aug 13, 2024)"
}

@InProceedings{nr_Dingetal2024,
    author="Leyuan Ding et. al.",
    editor="Arai, Kohei",
    title="Can Hallucination Reduction in LLMs Improve Online Sexism Detection?",
    booktitle="Intelligent Systems and Applications",
    year="2024",
    publisher="Springer Nature Switzerland",
    pages="625--638",
}

@online{nr_Google2024,
    author = "Google",
    title = "Classification: Accuracy, recall, precision, and related metrics",
    url = "https://developers.google.com/machine-learning/crash-course/classification/accuracy-precision-recall",
    year = "2024",
    addendum = "(accessed: Jan 27, 2025)"
}

@online{nr_Zhaoetal2024,
    author = "Yibo Zhao et. al.",
    title = "Enhancing LLM-based Hatred and Toxicity Detection with Meta-Toxic Knowledge Graph",
    url = "https://arxiv.org/abs/2412.15268",
    year = "2024"
}

@online{nr_Pérezetal2024,
    author = "Juan Manuel Pérez et. al.",
    title = "Exploring Large Language Models for Hate Speech Detection in Rioplatense Spanish",
    url = "https://arxiv.org/abs/2410.12174",
    year = "2024"
}

@article{nr_Kang2024,
    author = "Jungho Kang",
    title = "Federated Learning and LLM-based Social Media Comment Classification System Using Crowdsourcing Techniques",
    journal = "International Journal of Computer Science and Network Security",
    volume = "24",
    page = "25-31"
}

@online{nr_Joetal2024,
    author = "Claire Wonjeong Jo et. al.",
    title = "Harmful YouTube Video Detection: A Taxonomy of Online Harm and MLLMs as Alternative Annotators",
    url = "https://arxiv.org/abs/2411.05854",
    year = "2024"
}

@online{nr_Ouyangetal2024,
    author = "Rongxin Ouyang et. al.",
    title = "Hateful Meme Detection through Context-Sensitive Prompting and Fine-Grained Labeling",
    url = "https://arxiv.org/abs/2411.10480",
    year = "2024"
}

@online{nr_Wangetal2024,
    author = "ZhentingWang et. al.",
    title = "MLLM-as-a-Judge for Image Safety without Human Labeling",
    url = "https://arxiv.org/abs/2501.00192",
    year = "2024"
}

@inproceedings{nr_Sasidaran_J2024,
    author = {Keerthana Sasidaran and Geetha J},
    year = {2024},
    pages = {1-6},
    title = {Multimodal Hate Speech Detection using Fine-Tuned Llama 2 Model},
    booktitle = "2024 International Conference on Intelligent Algorithms for Computational Intelligence Systems",
    doi = {10.1109/IACIS61494.2024.10722018}
}

@online{nr_Xuetal2024,
    author = "Yiwen Xu et. al.",
    title = "Safe Guard: an LLM-agent for Real-time Voice-based Hate Speech Detection in Social Virtual Reality",
    url = "https://arxiv.org/abs/2409.15623",
    year = "2024"
}

@online{nr_Antypasetal2024,
    author = "Dimosthenis Antypas et. al.",
    title = "Sensitive Content Classification in Social Media: A Holistic Resource and Evaluation",
    url = "https://arxiv.org/abs/2411.19832",
    year = "2024"
}

@online{nr_AlDahouletal2024,
    author = "Nouar AlDahoul et. al.",
    title = "Advancing Content Moderation: Evaluating Large Language Models for Detecting Sensitive Content Across Text, Images, and Videos",
    url = "https://arxiv.org/abs/2411.17123",
    year = "2024"
}

@online{nr_González-Bustamante2024,
    author = "Bastián González-Bustamante",
    title = "Benchmarking LLMs in Political Content Text-Annotation: Proof-of-Concept with Toxicity and Incivility Data",
    url = "https://arxiv.org/abs/2409.09741",
    year = "2024"
}

@article{nr_Jhaveretal2019,
    author = "Shagun Jhaver et. al.",
    title = "'Did You Suspect the Post Would be Removed?': Understanding User Reactions to Content Removals on Reddit",
    year = "2019",
    volume = "3",
    url = "https://doi.org/10.1145/3359294",
    doi = "10.1145/3359294",
    journal = "Proc. ACM Hum.-Comput. Interact.",
    numpages = "33",
}

@article{nr_Latifetal2025,
  author = "Siddique Latif et. al.",
  journal = "IEEE Computational Intelligence Magazine", 
  title= "Can Large Language Models Aid in Annotating Speech Emotional Data? Uncovering New Frontiers", 
  year = "2025",
  volume = "20",
  number = "1",
  pages = "66-77",
  doi = "10.1109/MCI.2024.3504833"
}

@online{nr_Meta2020,
    author = "Meta",
    title = "How Facebook uses super-efficient AI models to detect hate speech",
    url = "https://ai.meta.com/blog/how-facebook-uses-super-efficient-ai-models-to-detect-hate-speech/",
    year = "2020",
    addendum = "(accessed: Jan 27, 2025)"
}

@article{nr_Penagos2024,
    author = "Emmanuel Vargas Penagos",
    title = "ChatGPT, can you solve the content moderation dilemma?",
    journal = "International Journal of Law and Information Technology",
    volume = "32",
    number = "1",
    pages = "1-18",
    year = "2024",
    doi = "10.1093/ijlit/eaae028"
}

@online{nr_Lietal2024,
    author = "Yaqiong Li et. al.",
    title = "DeMod: A Holistic Tool with Explainable Detection and Personalized Modification for Toxicity Censorship",
    url = "https://arxiv.org/abs/2411.01844",
    year = "2024"
}

@article{nr_Gielensetal2025,
    author = "Erwin Gielens et. al.",
    title = "Goodbye human annotators? Content analysis of social policy debates using ChatGPT",
    journal = "Journal of Social Policy",
    pages = "1-20",
    year = "2025",
    doi = "10.1017/S0047279424000382"
}

@article{nr_Ruckenstein_Turunen2020,
    author = "Minna Ruckenstein and Linda Lisa Maria Turunen",
    title = "Re-humanizing the platform: Content moderators and the logic of care",
    journal = "new media \& society",
    pages = "1026–1042",
    volume = "22",
    year = "2020",
    doi = "10.1177/1461444819875990"
}

@inproceedings{nr_Caoetal2024,
    title = "Toxicity Detection is {NOT} all you Need: Measuring the Gaps to Supporting Volunteer Content Moderators through a User-Centric Method",
    author = "Yang Trista Cao et. al.",
    booktitle = "Proceedings of the 2024 Conference on Empirical Methods in Natural Language Processing",
    year = "2024",
    address = "Miami, Florida, USA",
    doi = "10.18653/v1/2024.emnlp-main.209",
    pages = "3567--3587",
}

@online{nr_Songetal2025,
    author = "Jiaxin Song et. al.",
    title = "U-GIFT: Uncertainty-Guided Firewall for Toxic Speech in Few-Shot Scenario",
    url = "https://arxiv.org/abs/2501.00907",
    year = "2025"
}

@ARTICLE{nr_Kumaretal2024,
  author={Raghvendra Kumar et. al.},
  journal={IEEE Transactions on Artificial Intelligence}, 
  title={Silver Lining in the Fake News Cloud: Can Large Language Models Help Detect Misinformation?}, 
  year={2025},
  volume={6},
  number={1},
  pages={14-24},
  }

@online{nr_Villate-Castilloetal2024,
    author = "Guillermo Villate-Castillo et. al.",
    title = "A Collaborative Content Moderation Framework for Toxicity Detection based on Conformalized Estimates of Annotation Disagreement",
    url = "https://arxiv.org/abs/2411.04090v2",
    year = "2024"
}

@online{nr_Guerdanetal2024,
    author = "Luke Guerdan et. al.",
    title = "A Framework for Evaluating LLMs Under Task Indeterminacy",
    url = "https://arxiv.org/abs/2411.13760",
    year = "2024"
}

@inproceedings{nr_Chen_Mueller2024,
    title = "Quantifying Uncertainty in Answers from any Language Model and Enhancing their Trustworthiness",
    author = "Jiuhai Chen  and Jonas Mueller",
    booktitle = "Proceedings of the 62nd Annual Meeting of the Association for Computational Linguistics (Volume 1: Long Papers)",
    year = "2024",
    publisher = "Association for Computational Linguistics",
    url = "https://aclanthology.org/2024.acl-long.283/",
    doi = "10.18653/v1/2024.acl-long.283",
    pages = "5186--5200",
}

@inproceedings{nr_Leeetal2023,
    title = "Can Large Language Models Capture Dissenting Human Voices?",
    author = "Noah Lee et. al.",
    booktitle = "Proceedings of the 2023 Conference on Empirical Methods in Natural Language Processing",
    year = "2023",
    publisher = "Association for Computational Linguistics",
    url = "https://aclanthology.org/2023.emnlp-main.278/",
    doi = "10.18653/v1/2023.emnlp-main.278",
    pages = "4569--4585",
}

@online{nr_Xiongetal2023,
    author = "Miao Xiong et. al.",
    title = "Can LLMs Express Their Uncertainty? An Empirical Evaluation of Confidence Elicitation in LLMs",
    url = "https://arxiv.org/abs/2306.13063",
    year = "2023"
}

@online{nr_Shorinwaetal2024,
    author = "Ola Shorinwa et. al.",
    title = "A Survey on Uncertainty Quantification of Large Language Models: Taxonomy, Open Research Challenges, and Future Directions",
    url = "https://arxiv.org/abs/2412.05563",
    year = "2024"
}

@online{nr_Stengel-Eskinetal2024,
    author = "Elias Stengel-Eskin et. al.",
    title = "LACIE: Listener-Aware Finetuning for Confidence Calibration in Large Language Models",
    url = "https://arxiv.org/abs/2405.21028",
    year = "2024"
}

@online{nr_Becker_Soatto2024,
    author = "Evan Becker and Stefano Soatto",
    title = "Cycles of Thought: Measuring LLM Confidence through Stable Explanations",
    url = "https://arxiv.org/abs/2406.03441",
    year = "2024"
}

@online{nr_Yangetal2024,
    author = "Daniel Yang et. al.",
    title = "On Verbalized Confidence Scores for LLMs",
    url = "https://arxiv.org/abs/2412.14737",
    year = "2024"
}

@article{nr_Bodaghietal2024,
author = "Arezo Bodaghi et. al.",
title = "AugmenToxic: Leveraging Reinforcement Learning to Optimize LLM Instruction Fine-Tuning for Data Augmentation to Enhance Toxicity Detection",
year = "2024",
publisher = "Association for Computing Machinery",
address = "New York, NY, USA",
doi = "10.1145/3700791",
journal = "ACM Trans. Web",
}

@online{nr_Parketal2024,
    author = "Jinkyung Katie Park et. al.",
    title = "Collaborative Human-AI Risk Annotation: Co-Annotating Online Incivility with CHAIRA",
    url = "https://arxiv.org/abs/2409.14223",
    year = "2024"
}

@inproceedings{nr_Wangetal2023,
author = "Han Wang et. al.",
title = "Evaluating GPT-3 generated explanations for hateful content moderation",
year = "2023",
doi = "10.24963/ijcai.2023/694",
booktitle = "Proceedings of the Thirty-Second International Joint Conference on Artificial Intelligence",
series = {IJCAI '23}
}

@inproceedings{nr_Linetal2024,
author = "Hongzhan Lin et. al.",
title = "Towards Explainable Harmful Meme Detection through Multimodal Debate between Large Language Models",
year = "2024",
address = "New York, NY, USA",
doi = "10.1145/3589334.3645381",
booktitle = "Proceedings of the ACM Web Conference 2024",
pages = "2359–2370",
}

@online{nr_Kroegeretal2024,
    author = "Nicholas Kroeger et. al.",
    title = "Are Large Language Models Post Hoc Explainers?",
    url = "https://arxiv.org/abs/2310.05797v3",
    year = "2024"
}

@online{nr_Weietal2024,
    author = "Jerry Wei et. al.",
    title = "Simple synthetic data reduces sycophancy in large language models",
    url = "https://arxiv.org/abs/2308.03958",
    year = "2024"
}

@inproceedings{nr_Bonaventuraetal2025,
    title = "From Detection to Explanation: Effective Learning Strategies for LLMs in Online Abusive Language Research",
    author = "Chiara Di Bonaventura et. al.",
    booktitle = "Proceedings of the 31st International Conference on Computational Linguistics",
    year = "2025",
    publisher = "Association for Computational Linguistics",
    url = "https://aclanthology.org/2025.coling-main.141/",
    pages = "2067--2084"
}

@inproceedings{nr_Dönmezetal2024,
    title = "Please note that I'm just an AI: Analysis of Behavior Patterns of LLMs in (Non-)offensive Speech Identification",
    author = "Esra Dönmez et. al.",
    booktitle = "Proceedings of the 2024 Conference on Empirical Methods in Natural Language Processing",
    year = "2024",
    publisher = "Association for Computational Linguistics",
    doi = "10.18653/v1/2024.emnlp-main.1019",
    pages = "18340--18357"
}

@online{nr_Nirmal2024,
    author = "Ayushi Nirmal",
    title = "Interpretable Hate Speech Detection via Large Language Model-extracted Rationales",
    url = "https://keep.lib.asu.edu/items/193452",
    year = "2024"
}

@inproceedings{nr_Bonaventuraetal2024,
title = "Is Explanation All You Need? An Expert Survey on LLM-generated Explanations for Abusive Language Detection",
author = "Chiara Di Bonaventura et. al.",
year = "2024",
url = "https://kclpure.kcl.ac.uk/portal/en/publications/is-explanation-all-you-need-an-expert-survey-on-llm-generated-exp",
booktitle = "Tenth Italian Conference on Computational Linguistics (CLiC-it 2024)"
}

@inproceedings{nr_Kunz_Kuhlmann2024,
    title = "Properties and Challenges of LLM-Generated Explanations",
    author = "Jenny Kunz and Marco Kuhlmann",
    booktitle = "Proceedings of the Third Workshop on Bridging Human--Computer Interaction and Natural Language Processing",
    year = "2024",
    publisher = "Association for Computational Linguistics",
    url = "https://aclanthology.org/2024.hcinlp-1.2/",
    doi = "10.18653/v1/2024.hcinlp-1.2",
    pages = "13--27"
}

@inproceedings{nr_Cabelloetal2023,
    title = "On the Independence of Association Bias and Empirical Fairness in Language Models",
    author = "Laura Cabello et. al.",
    booktitle = "Proceedings of the 2023 ACM Conf. on Fairness, Accountability, and Transparency",
    year = "2023",
    doi = "10.1145/3593013.3594004",
    pages = "370-378"
}

@article{nr_Fergusonetal2024,
author = "Sharon Ferguson et. al.",
title = "The Explanation That Hits Home: The Characteristics of Verbal Explanations That Affect Human Perception in Subjective Decision-Making",
year = "2024",
volume = "8",
doi = "10.1145/3687056",
journal = "Proc. ACM Hum.-Comput. Interact."
}

@online{nr_EuropeanUnion2022,
    author = "European Union",
    title = "Digital Services Act (Directive 2000/31/EC)",
    url = "https://eur-lex.europa.eu/legal-content/EN/TXT/?uri=CELEX%3A32022R2065",
    year = "2022"
}

@online{nr_Piot_Parapar2024,
    author = "Paloma Piot and Javier Parapar",
    title = "Towards Efficient and Explainable Hate Speech Detection via Model Distillation",
    url = "https://arxiv.org/abs/2412.13698",
    year = "2024"
}

@online{nr_Singhetal2024,
    author = "Chandan Singh et. al.",
    title = "Rethinking Interpretability in the Era of Large Language Models",
    url = "https://arxiv.org/abs/2402.01761",
    year = "2024"
}

@inproceedings{nr_Vaccaroetal2021,
author = "Kristen Vaccaro et. al.",
title = "Contestability For Content Moderation",
booktitle = "Proceedings of the ACM on Human-Computer Interaction",
year = "2021",
publisher = "Association for Computing Machinery",
volume = "5",
doi = "10.1145/3476059"
}

@online{nr_Heetal2024,
    author = "Jerry Zhi-Yang He et. al.",
    title = "CoS: Enhancing Personalization and Mitigating Bias with Context Steering",
    url = "https://arxiv.org/abs/2405.01768",
    year = "2024"
}

@article{nr_Chaoetal2024,
    author = "August F.Y. Chao et. al.",
    title = "From hate to harmony: Leveraging large language models for safer speech in times of COVID-19 crisis",
    journal = "Heliyon",
    volume = "10",
    year = "2024",
    doi = "10.1016/j.heliyon.2024.e35468"
}

@inproceedings{nr_Ziegenbeinetal2024,
    title = "LLM-based Rewriting of Inappropriate Argumentation using Reinforcement Learning from Machine Feedback",
    author = "Timon Ziegenbein et. al.",
    booktitle = "Proceedings of the 62nd Annual Meeting of the Association for Computational Linguistics (Volume 1: Long Papers)",
    year = "2024",
    publisher = "Association for Computational Linguistics",
    doi = "10.18653/v1/2024.acl-long.244",
    pages = "4455--4476",
}

@inproceedings{nr_Gabrieletal2024,
    title = "MisinfoEval: Generative AI in the Era of 'Alternative Facts'",
    author = "Saadia Gabriel et. al.",
    booktitle = "Proceedings of the 2024 Conference on Empirical Methods in Natural Language Processing",
    year = "2024",
    publisher = "Association for Computational Linguistics",
    doi = "10.18653/v1/2024.emnlp-main.487",
    pages = "8566--8578",
}

@article{nr_Elkin-Koren_Perel2023,
    author = "Niva Elkin-Koren and Maayan Perel",
    title = "Speech Contestation by Design: Democratizing Speech Governance by AI",
    journal = "Florida State University Law Review",
    volume = "50",
    pages = "611-665",
    year = "2023"
}

@article{nr_Nahmias_Perel2021,
    author = "Yifat Nahmias and Maayan Perel",
    title = "The Oversight of Content Moderation by AI: Impact Assessments and Their Limitations",
    journal = "Harvard Journal on Legislation",
    volume = "58",
    pages = "145-194",
    year = "2021"
}

@inproceedings{nr_Masudetal2024,
    title = "Hate Personified: Investigating the role of {LLM}s in content moderation",
    author = "Sarah Masud et. al.",
    booktitle = "Proceedings of the 2024 Conference on Empirical Methods in Natural Language Processing",
    year = "2024",
    publisher = "Association for Computational Linguistics",
    doi = "10.18653/v1/2024.emnlp-main.886",
    pages = "15847--15863",
}

@online{nr_YuxinWangetal2024,
    author = "Yuxin Wang et. al.",
    title = "ImpScore: A Learnable Metric For Quantifying The Implicitness Level of Language",
    url = "https://arxiv.org/abs/2411.05172v2",
    year = "2024"
}

@online{nr_Roumeliotisetal2024,
    author = "Konstantinos I. Roumeliotis et. al.",
    title = "Unmasking Misinformation: Leveraging CNN, BERT, and GPT Models for Robust Fake News Classification",
    url = "https://doi.org/10.20944/preprints202411.0786.v1",
    year = "2024"
}

@inproceedings{nr_Mathewetal2021,
    author = "Binny Mathew et. al.",
    title = "HateXplain: A Benchmark Dataset for Explainable Hate Speech Detection",
    booktitle = "The Thirty-Fifth AAAI Conference on Artificial Intelligence",
    year = "2021",
    publisher = "Association for the Advancement of Artificial Intelligence",
    pages = "14867-14875"
}

@article{nr_Maliketal2024,
    author = "Jitendra Singh Malik et. al.",
    title = "Deep learning for hate speech detection: a comparative study",
    journal = " International Journal of Data Science and Analytics",
    doi = "10.1007/s41060-024-00650-6",
    year = "2024"
}

@inproceedings{nr_YizhuoZhangetal2024,
    author = "Yizhuo Zhang et. al.",
    title = "Can LLM Graph Reasoning Generalize beyond Pattern Memorization?",
    booktitle = "Findings of the Association for Computational Linguistics: EMNLP",
    pages = " 2289–2305",
    year = "2024"
}

@online{nr_ChenghaoZhuetal2024,
    author = "Chenghao Zhu et. al.",
    title = "Is Your LLM Outdated? Evaluating LLMs at Temporal Generalization",
    url = "https://arxiv.org/abs/2405.08460",
    year = "2024"
}

@online{nr_Sharmaetal2023,
    author = "Mrinank Sharma et. al.",
    title = "Towards Understanding Sycophancy in Language Models",
    url = "https://arxiv.org/abs/2310.13548",
    year = "2023"
}

@article{nr_LeiHuangetal2025,
    author = "Lei Huang et. al.",
    title = "A Survey on Hallucination in Large Language Models: Principles, Taxonomy, Challenges, and Open Questions",
    year = "2025",
    publisher = "Association for Computing Machinery",
    volume = "43",
    doi = "10.1145/3703155",
    journal = "ACM Trans. Inf. Syst.",
}

@article{nr_Gallegosetal2024,
    author = "Isabel O. Gallegos et. al.",
    title = "Bias and Fairness in Large Language Models: A Survey",
    journal = "Computational Linguistics",
    volume = "50",
    pages = "1097-1179",
    year = "2024",
    doi = "10.1162/coli_a_00524",
}

@online{nr_ZiweiXuetal2024,
    author = "Ziwei Xu et. al.",
    title = "Hallucination is Inevitable: An Innate Limitation of Large Language Models",
    url = "https://arxiv.org/abs/2401.11817",
    year = "2024"
}

@inproceedings{nr_Manvietal2024,
author = "Rohin Manvi et. al.",
title = "Large Language Models are Geographically Biased",
year = "2024",
booktitle = "Proceedings of the 41st International Conference on Machine Learning",
location = "Vienna, Austria",
series = "ICML'24"
}

@article{nr_Motokietal2024,
    author = "Fabio Motoki et. al.",
    title = "More human than human: measuring ChatGPT political bias",
    journal = "Public Choice",
    volume = "198",
    pages = "3-23",
    year = "2024",
    doi = "10.1007/s11127-023-01097-2"
}

@online{nr_Dasetal2024,
    author = "Amit Das et. al.",
    title = "Investigating Annotator Bias in Large Language Models for Hate Speech Detection",
    url = "https://arxiv.org/abs/2406.11109",
    year = "2024"
}

@online{nr_YueZhangetal2023,
    author = "Yue Zhang et. al.",
    title = "Siren's Song in the AI Ocean: A Survey on Hallucination in Large Language Models",
    url = "https://arxiv.org/abs/2309.01219",
    year = "2023"
}

@inproceedings{nr_JunyiLietal2024,
    title = "The Dawn After the Dark: An Empirical Study on Factuality Hallucination in Large Language Models",
    author = "Junyi Li et. al.",
    booktitle = "Proceedings of the 62nd Annual Meeting of the Association for Computational Linguistics (Volume 1: Long Papers)",
    year = "2024",
    publisher = "Association for Computational Linguistics",
    doi = "10.18653/v1/2024.acl-long.586",
    pages = "10879--10899",
}

@online{nr_Levinetal2016,
    author = "Sam Levin et. al.",
    title = "Facebook backs down from 'napalm girl' censorship and reinstates photo",
    publisher = "The Guardian",
    url = "https://www.theguardian.com/technology/2016/sep/09/facebook-reinstates-napalm-girl-photo",
    year = "2016"
}

@online{FacebookOversightBoard2021,
    author    = "Facebook Oversight Board",
    title     = "PRO-NAVALNY PROTESTS IN RUSSIA",
    url       = "https://www.oversightboard.com/decision/fb-6yhrxhzr/",
    year = "2021",
    addendum = "(accessed: Feb 3, 2025)"
}

@book{nr_Gwet2014,
    author = "Kilem L Gwet",
    title = "Handbook of Inter-Rater Reliability: The Definitive Guide to Measuring the Extent of Agreement Among Raters",
    publisher = "Advanced Analytics",
    year = "2014"
}

@misc{Marshall1980,
    author = "Thurgood Marshall", 
    title = "Marshall v. Jerrico, Inc., 446 U.S. 238",
    page = "242",
    year = "1980",
    url = "https://supreme.justia.com/cases/federal/us/446/238/"
}

@misc{AccessNowetal2021,
    author = "Access Now et. al.",
    title = "The Santa Clara Principles",
    year = "2021",
    url = "https://santaclaraprinciples.org/"
}

\end{document}